\documentclass[journal]{IEEEtran}
\usepackage{cite}

\ifCLASSINFOpdf
\else
\fi
\usepackage{algorithm}
\usepackage{algpseudocode}
\usepackage{array} 
\ifCLASSOPTIONcompsoc\usepackage[caption=false,font=normalsize,labelfont=sf,textfont=sf]{subfig}
\else
\usepackage[caption=false,font=footnotesize]{subfig}
\fi
\usepackage{xcolor}
\usepackage{graphicx}
\usepackage{amsfonts}

\usepackage{breqn}
\begin{document}
%
\title{LiDAL-Assisted RLNC-NOMA in OWC Systems}
%
%
%

\author{Ahmed A. Hassan,~\IEEEmembership{Member,~IEEE,}
       Ahmad Adnan Qidan,~\IEEEmembership{Member, IEEE,}
       Taisir Elgorashi,~\IEEEmembership{Member, IEEE,}
        and~Jaafar Elmirghani,~\IEEEmembership{Fellow,~IEEE}
}
%
%

\markboth{Journal of \LaTeX\ Class Files,~Vol.~14, No.~8, August~2025}%
{Shell \MakeLowercase{\textit{et al.}}: Bare Demo of IEEEtran.cls for IEEE Journals}
%



\maketitle

\begin{abstract} 
Optical wireless communication (OWC) is a key enabler for immersive indoor data transmission in future wireless networks. However, managing multi-user interference remains challenging in dense indoor OWC deployments with multiple optical access points (APs) serving multiple users.
In this paper, we propose a dual-function OWC system for communication and localization. For data transmission, we design non-orthogonal multiple access (NOMA) combined with random linear network coding (RLNC): NOMA serves multiple users simultaneously via power-domain multiplexing, while RLNC improves decoding robustness under residual noise and processing impairments. The system is assisted by a light detection and localization (LiDAL) module that estimates users' spatio-temporal presence and location to enable dynamic user association. LiDAL also supports channel state information (CSI) estimation in realistic indoor scenarios where CSI may be noisy and/or highly correlated. We evaluate RLNC-NOMA by analyzing the probability of successful decoding and comparing it with conventional NOMA and orthogonal schemes. In addition, we derive the Cram\'er--Rao lower bound (CRLB) to evaluate the localization accuracy. The results indicate that RLNC-NOMA increases the probability of successful decoding, and that the proposed unbiased location estimator achieves 44\% better CSI accuracy than the baseline CSI estimation and improves the overall system performance.
\end{abstract}
%
\IEEEpeerreviewmaketitle

\section{Introduction}
%
%
%
%

\IEEEPARstart{O}{}{ptical wireless communication (OWC) has emerged as a key technology for the next generation of indoor cellular systems (i.e., 6G), aiming to meet the increasing demands of massive device connectivity and ultra-low latency data services such as augmented and virtual reality (AR/VR) and real-time gaming}\cite{alsulami_optical_nodate}. OWC has received great interest from both academia and industry as it can enable multi-gigabit transmissions to indoor users by utilizing free-license optical spectrum and reusing existing illumination infrastructure\cite{9521837},\cite{9064520}. Furthermore, {OWC systems can use different wavelengths (i.e., colors) to send information to multiple users}, which would provide high spectral efficiency and security. In a typical OWC, the optical source, such as light-emitting diode (LED) or laser diode (LD) converts the data symbols into a form of optical signal, and the receiver equipped with a photo-diode detector (PD) converts the received optical power into electrical current that represents to the transmitted data symbols\cite{pathak_visible_2015}. 

To satisfy the illumination requirements of a small indoor area, several optical sources (i.e., LEDs or LDs) are required. As a result, OWC systems would notably suffer from high levels of interference, especially in densely multi-user indoor settings. Such interference poses a considerable degradation in the spectrum efficiency and strongly dictates the achievable user data rates within such scenarios. For this reason, managing user medium access becomes unavoidable\cite{8636954,9500371}.

In this context, non-orthogonal multiple access (NOMA) has recently garnered widespread attention as a promising multiple access (MA) candidate for indoor OWC systems. Compared to conventional orthogonal multiple access (OMA) such as (i.e., TDMA, OFDMA, CDMA,\dots,etc.), NOMA would significantly improve spectral efficiency and provide a balanced throughput and fairness trade-off\cite{feng_multiple_2019}. Typical NOMA controls the power domain (PD-NOMA) to serve multiple users at different power levels of the same channel resource by applying the super positioning coding (SC) technique on transmitted users signals\cite{zhang_user_2017}. At the receiving end, successive interference cancellation (SIC) is employed to decode the signal of the intended user.

NOMA is applicable to uplink and downlink OWC transmissions. In downlink NOMA, users with weak channel gains are usually assigned more optical power than users with strong channel gains. In this case, users with weak channel gain can decode their signals while treating the signals of others as noise. The performance of NOMA in OWC systems has been thoroughly investigated in various indoor environments and configurations \cite{ yin_performance_2015, marshoud_performance_2017}. In which NOMA has shown that it can outperform traditional OMA in several key metrics such as the total ergodic sum rate and the outage probability. Like any technology, NOMA suffers from high data packet loss due to the imperfection of the SIC that results from residual noise during the separation process of the corresponding user signal.

Random linear network coding (RLNC) is introduced in \cite{ho_random_2006} as a way to enhance transmission efficiency by combining data pieces (i.e., packets) with random coefficients, which can be independently chosen from a finite Galois field (GF). In RLNC, the coded packets can be transmitted instead of original packets, then users decode the corresponding original data packets using, for instance, a Gauss-Jordan elimination process. In fact, RLNC can efficiently improve system capacity and reduce the number of retransmissions. Moreover, RLNC does not require any code sequence synchronization and/or packet loss feedback \cite{lin_efficient_2014} increasing robustness in various cluttered wireless environments. In OWC systems, several studies have proposed network coding to enhance transmission capacity, such as physical-layer network coding (PNC) \cite{hong_adaptive_2017} and relay-based network coding \cite{wang_energy-efficient_2018},\cite{hong_channel-aware_2020}. For NOMA systems, RLNC is proposed to improve the packet success probability and the packet delay in RF- and OWC-based multicast networks as reported in \cite{park_random_2015} and \cite{hassan_random_2023-1}, respectively.

In the last decade, most of the research in indoor NOMA-based systems has considered channel state information (CSI) an intuitive parameter for user association and ranking, power allocation, and signal encoding/decoding.  CSI has been evaluated under the assumption that the wireless channel conditions remain constant for a given time unit or varies independently following a composition of probabilistic models, i.e. Rayleigh and exponential distributions\cite{marshoud_performance_2017}, \cite{yang_performance_2016}. However, utilizing location-based information in NOMA-based indoor systems did not receive the same attention as the CSI parameter, and that can be due to three main factors: low RF-based indoor positioning accuracy, induced signal processing overhead, and increased integration complexity\cite{zekavat_wireless_2011}. Therefore, adopting native object detection and localization capabilities within OWC systems and incorporating prior knowledge of user location information would improve NOMA-based OWC system efficiency. This approach enables optimized user access to resources and minimizes system design overhead, especially in indoor scenarios characterized by rapid channel fluctuation and high levels of interference\cite{zhou_visible_2023}.

Various studies considered enabling the visible light positioning (VLP) in OWC systems. The majority of the VLP techniques and mathematical methods addressed in the literature can be categorized into four main types: RSS-based methods (i.e., trilateration, proximity, fingerprinting), time-based methods (i.e., TOA, TDOA, PDOA)\cite{chen_indoor_2020},\cite{luo_indoor_2017} and angulation methods (i.e., AOA, AOD),\cite{luo_indoor_2017},\cite{eroglu_aoa-based_2015} or hybrid methods\cite{keskin_direct_2018}. In addition, VLP systems can be classified according to receiver type: non-imaging receivers (i.e., PD and angle diversity receiver (ADR))\cite{yang_indoor_2013}, and imaging receivers (i.e., camera)\cite{guan_robot_2021}.

For object detection, several mechanisms for object detection and recognition are proposed, such as infrared (IR)\cite{chen_local_2014}, laser radar (LADAR)\cite{nissinen_high_2016}, and visible light radar (LiDAR)\cite{guan_robot_2021}. These mechanisms can coexist in OWC systems to enable various capabilities, such as adaptive blockage avoidance and efficient resource management. In addition to that, several modified optical modulation schemes, such as PSS-PPM\cite{wen_pulse_2023} and IM-OFDM\cite{hawkins_im-ofdm_2024}, have recently emerged to simultaneously accommodate both detection and communication functions in one system. 

The studies in \cite{wang_position_2013, lv_high_2017}, demonstrated that the aforementioned VLP techniques can achieve high localization accuracy (i.e., in centimetres) within ideal indoor OWC systems. However, limited research has considered both the detection and localization of indoor objects (i.e., furniture, surfaces,\dots, etc.)  as well as user-associated properties such as skin color, worn clothes, and nomadic behaviors in realistic indoor OWC systems. Therefore, a light detection and localization system (LiDAL) was proposed in \cite{al-hameed_lidal_2019} to effectively use optical signals that are reflected from opaque objects and are captured by a non-imaging PD receiver that monitors the change of light intensity in the space-time domain. The LiDAL system can passively obtain spatio-temporal indoor information of user presence and location. {This offers a privacy-native advantage compared to imagery-based sensing and localization systems.}

{In this work, we propose a dual-function OWC system that integrates communication and localization and differs from existing approaches in three key aspects. First, unlike prior OWC and NOMA-based systems, we integrate multi-user NOMA with RLNC to improve spectral efficiency while mitigating error propagation in the NOMA signal decoding process. Second, in contrast to the LiDAL reference design \cite{al-hameed_lidal_2019}, we use LiDAL outputs not only for user detection and dynamic counting but also to enhance the CSI quality for the proposed RLNC-NOMA stack in realistic indoor scenarios where CSI is noisy and/or highly correlated. Third, we derive and analyze an unbiased location error estimator for the MIMO-based LiDAL architecture, which was not addressed in \cite{al-hameed_lidal_2019}. Finally, we further evaluate the performance of RLNC-NOMA by analyzing the probability of successful decoding and comparing it with conventional NOMA and OMA methods, and we evaluate LiDAL localization performance and its estimator precision in the considered indoor environment.
}

{This paper focuses on the design and operation of the LiDAL-integrated RLNC-NOMA system: we optimize localization accuracy and characterize location error for user association modeling. In our prior work \cite{hassan_drl-based_2026-1}, we instead optimized power for average sum rate and user fairness and investigated real-time machine-learning-based optimization.}

The article is organized as follows. In Section II, we introduce the proposed system model which describes in detail the components of the RLNC-NOMA and MIMO-LiDAL systems. Section III, we analyze the system design and derive the unbiased location error estimator. Additionally, the LiDAL-assisted user association scheme is formulated using the user location information obtained. Section IV presents the simulation results and performance evaluation. Finally, Section V concludes our findings and remarks.

\begin{figure*}[h!tbp]
        \centering
        \includegraphics[width=0.65\textwidth]{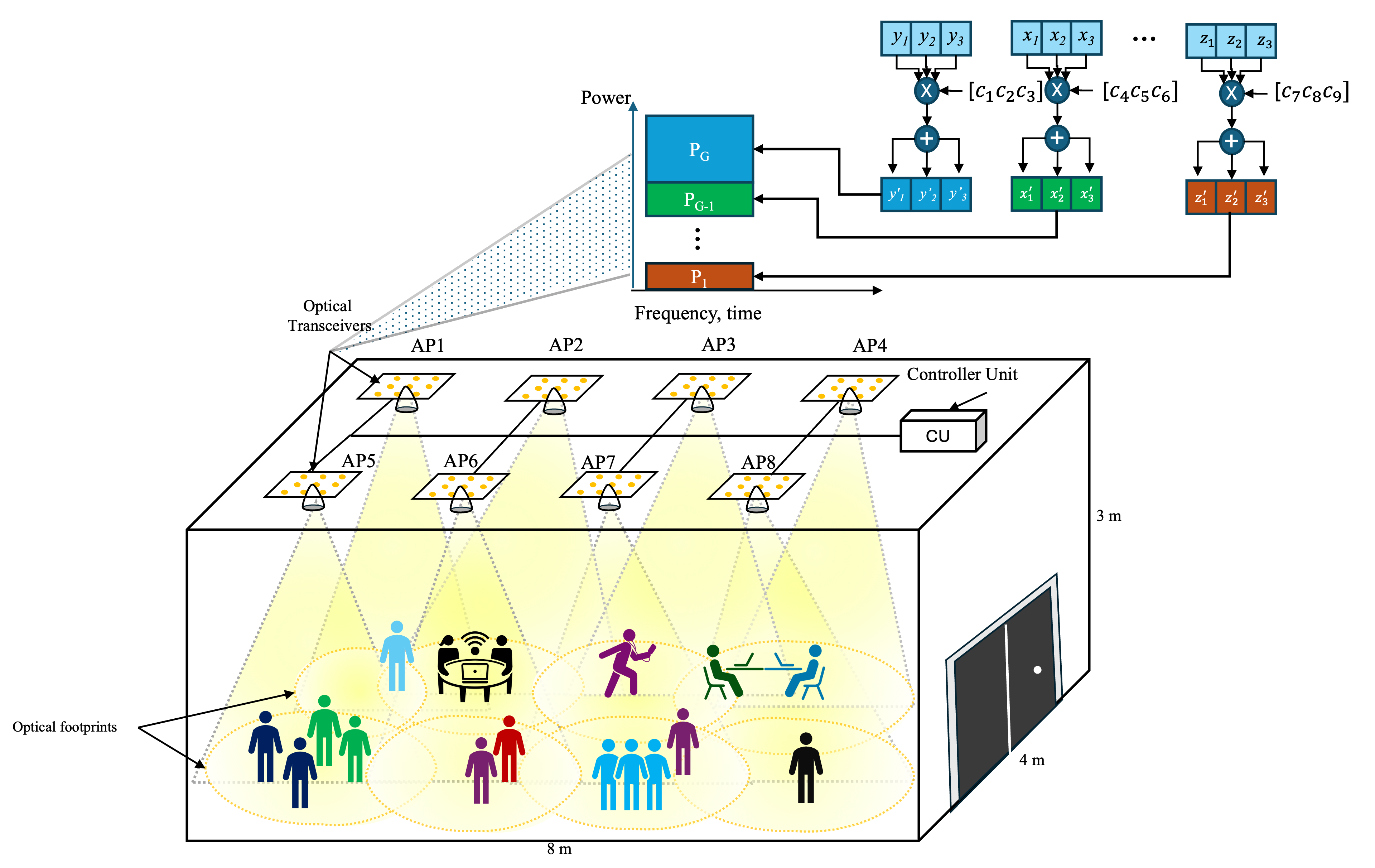}
        \caption{LiDAL-assisted RLNC-NOMA system model}
        \label{fig:system-model}
\end{figure*}

\section{System Model}
We consider an indoor OWC system model as illustrated in Fig.\ref{fig:system-model}, where multiple optical APs (i.e., transceivers) are installed on the ceiling to serve a large number of users in the communication plane.   This system model combines two systems: RLNC-NOMA system and MIMO-LiDAL system, both are linked to a central control unit (i.e., controller). The description of both system models and their components is discussed in detail below.
\subsection{ RLNC-NOMA System}
A multi-user multiple input single output (MU-MISO) downlink configuration is considered. {Each AP has a dual function transmitter \({Tx}_k^L\), $k = 1,2,\cdots,K$, (i.e., \(k\) optical footprint) that is equipped with a uniform array of homogenized RGB-LDs using in-front diffuser that emits a Lambertian radiation pattern and is utilized for both communication and localization as shown Fig.\ref{fig:system-model}}. A group-based NOMA is adopted to serve \(N\) communication users that are located and divided into at most \(K\) groups. Each user \(i\) (i.e., \(i\in M_k\)) associated with a transmitter \({Tx}_k^L\) is equipped with a single PD that decodes the information symbols received by that user. For simplicity, we assume that the OWC receivers of all system users are perpendicular to the communication plane and pointing toward the ceiling.
\subsubsection{RLNC-NOMA Channel Model}
In downlink NOMA scenario, \({Tx}_k^L\) simultaneously transmits real and non-negative superimposed signals \(\mathcal{S}_k\) with different power levels to convey information to \( M_k\) users of the associated group \(k\). Without loss of generality, the composed \(\mathcal{S}_k\) optical signals sent by optical AP \(k\) can be expressed by
\begin{align}
    \mathcal{S}_k = \mathcal{P}_t\sum_{i=1}^{M_k} \alpha_i s_i + I_{dc} \quad \forall i \in M_k, k \in K,
\end{align}
where \(s_i\) is the superimposed information signals and \(\alpha_i\) denotes the power allocation factor for user \(i\). \(\mathcal{P}_t\) denotes the total transmission power of an optical AP \(k\), and \(I_{dc}\) is the direct current bias to ensure the positive intensity of the transmitted signals. To maintain non-negativity and control the eye safety of the transmitted optical signal, the following constraints should be satisfied, respectively.
\begin{align}
    \mathcal{P}_t \sum^{M_k}_{i=1}\alpha_i\leq I_{dc},
    \label{eq:DC_current}
\end{align}
and
\begin{align}
   \mathcal{P}_t  \sum^{M_k}_{i=1}\alpha_i\leq A - I_{dc},
    \label{eq:max_optical_power}
\end{align}
where \(A\) is the maximum optical intensity of the corresponding optical AP \(k\). For any PD type receiver, the signal received by user \(i\) from optical AP \(k\) is given by
\begin{align}
    \mathcal{Y}_{i} = \mathfrak{R}_i\hbar_i \mathcal{P}_t\sum_{j=1}^{M_k} \alpha_j s_j + n_i \quad \forall i \in M_k, k \in K,
    \label{eq:owc-received-signal}
\end{align}
where \( \mathfrak{R}_i \) is the responsivity of PD, \(\hbar_i\) denotes the LOS channel gain coefficient, \(n_i\) is additive white Gaussian noise (AWGN) with zero mean and variance of \(\sigma^2_t=(\sigma_{thr}^2 + \sigma_{shot}^2)\) due to the noise of the user receiver and the ambient noise at its location. 

{Under the assumption that each user's receiver points vertically toward the optical AP with narrow field-of-view (FOV) angle (i.e., \( \Psi_c = 40^o\)), we consider the LOS component (usually represents 90\% of the overall received power) and neglect the diffuse components that are received due to reflections from the room walls and any other objects}.

Therefore, the channel gain of the LOS component between the transmitter \({Tx}_k^L\) of arbitrary optical AP \(k\) and the PD receiver of user \(i\) that belongs to \(M_k\) group of users is given by
\begin{multline}
      \hbar_i = \\
    \begin{cases}
        \dfrac{(m+1)A_{PD}}{2\pi {\Lambda_i^k}^2}cos^m(\phi_i)cos(\psi_i)T_f(\psi_i)g_c(\psi_i) \quad 0\leq \psi_i \leq \Psi_c,\\
        0 \quad \! \text{otherwise},
    \end{cases} \\
\end{multline}
where \(\phi_i\) and \(\psi_i\) donates the irradiance and incidence angles between the transmitter \({Tx}_k^L\) and the PD receiver of user \(i\), respectively. \(\Lambda_i^k\) denotes the measured access distance between the location of user \(i\) and its associated optical AP \(k\) which corresponds in particular to the LOS distance between the PD receiver of user \(i\) and the transmitter \({Tx}_k^L\) of the optical AP \(k\), \(\Psi_c\) represents the FOV angle of the PD receiver and \(A_{PD}\) denotes its physical area, \(m = - \frac{ln2}{ln\,cos{{\Phi}_{1/2}}}\) is the Lambertian index of \({Tx}_k^L\) corresponding half-power semi-angle denoted by \(\Phi_{1/2}\). Note that, \(T_f(\psi_i)\) and \(g_c(\psi_i)\) denote the optical filter gain of the PD receiver and the concentrator gain, respectively.

In this work, we assume that users are sorted in ascending order according to their location orders from the associated AP \(k\) (i.e., \(\Lambda_1^k \leq \Lambda_2^k\leq \cdots \leq \Lambda^k_{M_k}\)), as will be explained in detail later. In this way, we have \(\alpha_1 \leq \alpha_2 \leq \cdots \leq \alpha_{M_k}\) that satisfy the following NOMA conditions
\begin{align}
    \begin{cases}
       {0 \leq\alpha_i\leq 1}, \\
       {\sum^{M_k}_{i=1} \alpha_i =1}
    \end{cases} \quad {\forall i \in M_k, k \in K}.
\end{align}
At the receiver side, the SIC decoder of user \(i\)  decodes and subtracts the intended signals to all users at location \(X_m\) and have lower location order to the optical AP \(k\), where \((i+1 \leq m \leq M_k)\). Moreover, the signals of all users with higher location order, where \((1 \leq m \leq i)\) are treated as noise.
When SIC is performed on the superimposed information encoded using RLNC, and the signals intended for other users are subtracted, the received signal by the user \(i\) can be expressed as
\begin{dmath}
     \mathcal{Y}^*_i = \sum_{n=1}^f c_{in}.b_{in} + n^*_i \quad {\forall i \in M_k, k \in K},
     \label{eq:RLNC-RX-signal}
\end{dmath}
where \(c_{in}\) denotes a random chosen coding coefficient from the finite Galois field GF\((2^8)\), \(b_{in}\) is the original packet of \(f\) packets transmitted over the user information signal \(s_i\), and \(n^*_i\) denotes a real-valued AWGN with zero mean and variance \(\sigma^2_*\) that represents the noise resulting from the imperfect SIC attributed to noisy and/or correlated CSI of user \(i\).

Most of the literature contributions on NOMA are based on perfect knowledge of CSI (i.e., \(\hbar_i\)) which is not ideal for designing practical MIMO NOMA-based systems. In other contributions, NOMA transmission with imperfect CSI (i.e., \(\hbar_i^*\)) is considered instead of acquiring full channel CSI \cite{marshoud_performance_2017}. Furthermore, in distance-based NOMA systems, the user order may be suboptimal due to location estimation error and impaired channel conditions. However, exploiting additional location-based information using an integrated sensing and localization system (i.e., LiDAL) would efficiently capture the channel characteristics of users in a cluttered environment. 
Therefore, the location \(X_i\) of a user \(i\)  can be characterized by its lower bound of location error \(B^L_i\) and its access distance \( \Lambda_{i}^k\) to the associated optical AP \(k\). The corresponding imperfect CSI estimated error \(( \Delta \hbar_i^*= \hbar_i - \hbar_i^*)\) of user \(i\) at location \(X_i\) can be approximated by \cite{ma_robust_2023}
\begin{multline}
    \Delta \hbar_i^* \simeq \Delta_c \bigg(\frac{\eta^{m+1}}{({\Lambda_i^k}^2 + {B^L_i}^2)^{\frac{m+3}{2}}} - \frac{\eta^{m+1}}{({\Lambda_i^k}^2)^{\frac{m+3}{2}}}\bigg) \\ \forall i \in M_k, k \in K,
    \label{eq:LiDAL-NOMA-CSI-approx}
\end{multline}
where \(\Delta_c = \frac{(m+1)A_{PD}}{2\pi}T_f(\psi_i)g_c(\psi_i)\), { \(\eta\) denotes the height difference between a user \(i\) and the ceiling of the room that is assumed to be fixed due to the limited height variation of all users in the indoor environment considered \cite{al-hameed_lidal_2019}}. It is clear from the above that the imperfect CSI error is a function of two parameters, the estimated location error \(B^L_i\) and the measured access distance \(\Lambda_i^k\) of a user \(i\) that assigned to the optical AP \(k\). In fact, the CSI estimation error is inversely proportional to the estimated location error. Consequently, the RLNC-NOMA signal received by the user \(i\) can be easily derived from (\ref{eq:RLNC-RX-signal}) and (\ref{eq:LiDAL-NOMA-CSI-approx}), and (\ref{eq:owc-received-signal}) can be rewritten in the following form.
\begin{multline}
   \mathcal{Y}_{i} = \mathfrak{R}_i (\hbar_i^* + \Delta \hbar_i^*) \mathcal{P}_t\sum_{n=1}^f \alpha_i  c_{in}.b_{in} + n^'_i  \\ \forall i \in M_k, k \in K,
\end{multline}

hence, \(n^'_i = n^*_i + n_i\).
\subsubsection{RLNC Encoder}

\begin{figure}[t]
    \centering
    \includegraphics[width=0.8\linewidth]{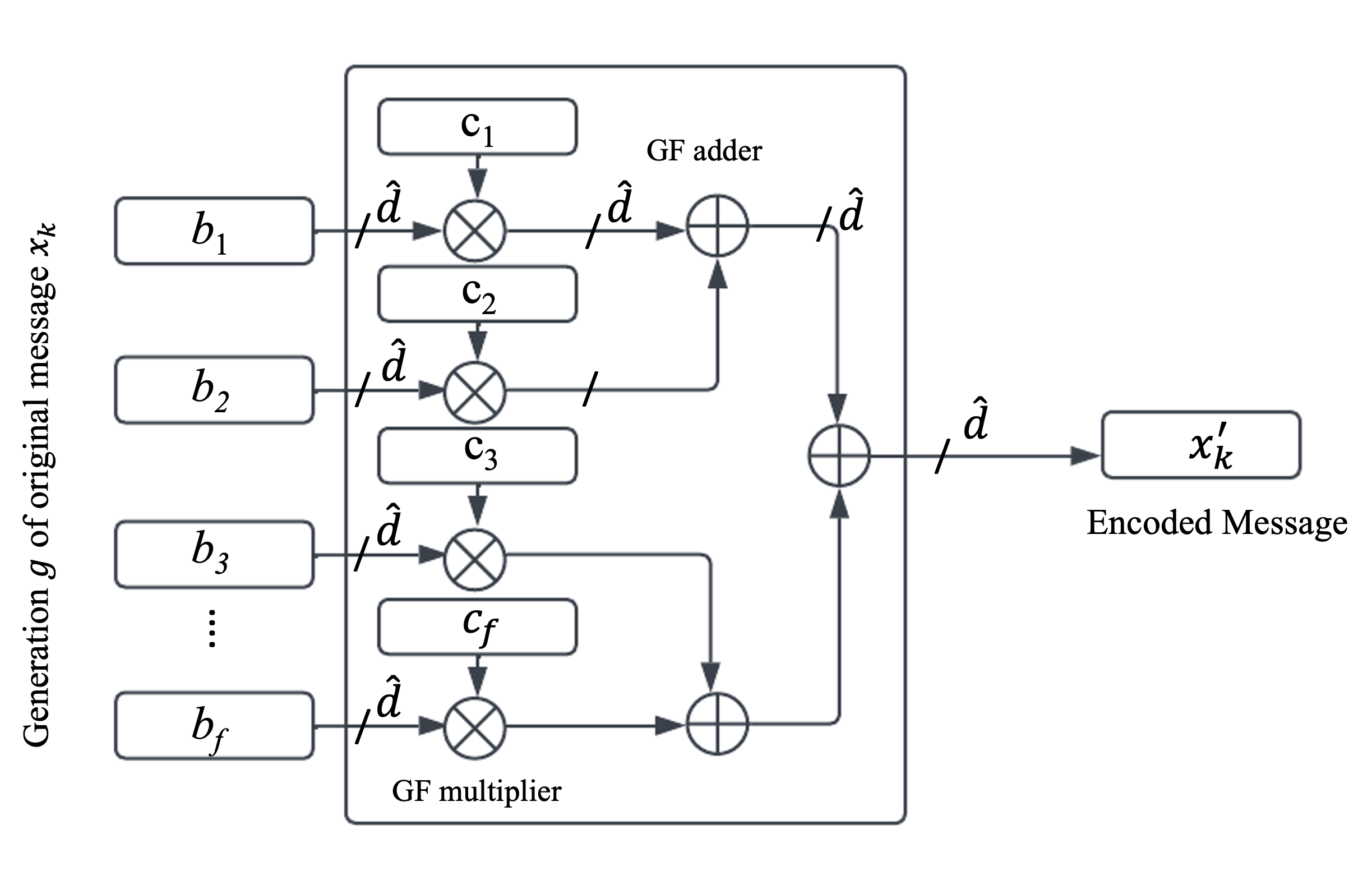}
    \caption{RLNC-NOMA encoder design.}
    \label{fig:RLNC-ENCODER}
\end{figure}

In an optical AP \(k\), the RLNC encoder is utilized to encode the message packets \(\mathbf{X}_k\) (original packets) in the proposed system model. We assume progressive encoding \cite{pedersen_kodo_2011} on a fixed-size buffer of original packets constituting a generation \(\textit{g}\), and each packet consists of a fixed length of \(\hat{d}\)-bytes.
Initially, a basic partitioning and interleaving procedure \cite{khan_characterisation_2018} generates a matrix of byte symbols that represents a single generation of packets. This matrix is then multiplied by a vector of coding coefficients randomly chosen from GF\((2^8)\). followed by a GF addition process to produce elements or symbols in the same finite field, forming the encoded message packets \(\mathbf{X}_k^\prime\) as illustrated in Fig.\ref{fig:RLNC-ENCODER}.
\subsubsection{RLNC Decoder}
\begin{figure}[th]
    \centering
    \includegraphics[width=0.8\linewidth]{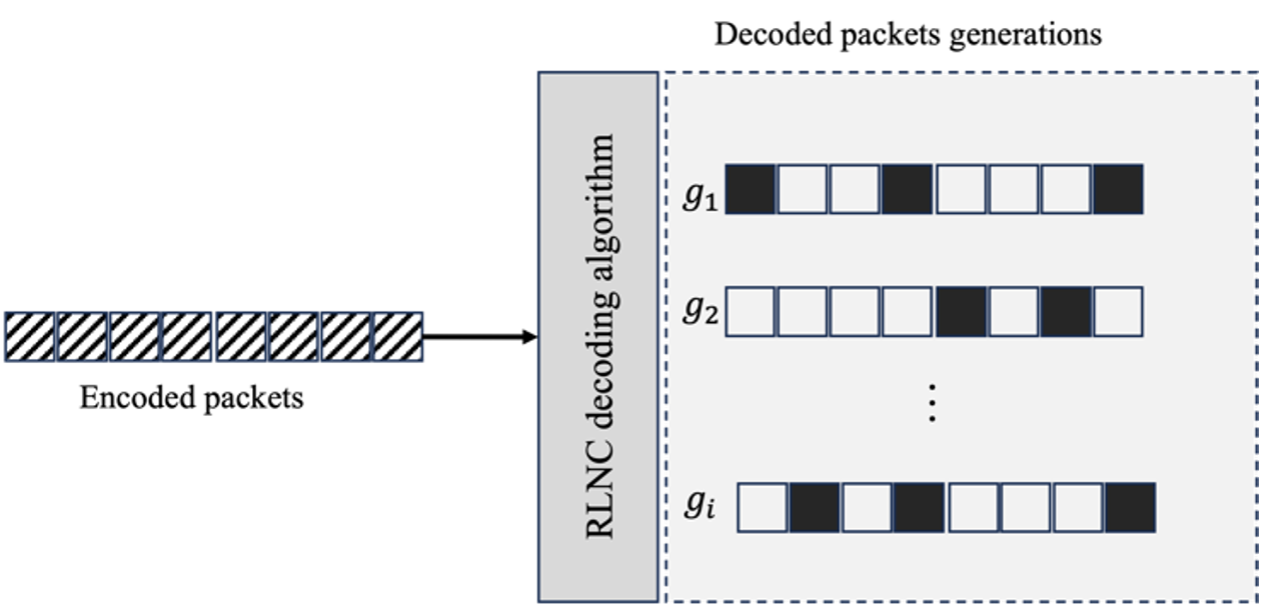}
    \caption{RLNC-NOMA decoder design.}
    \label{fig:RLNC-decoder}
\end{figure}

As soon as a sufficient number of linearly independent coded packets and the corresponding encoding coefficients are obtained after the optical signal separation process and demodulating the coded frames by the optical receiver of user 
\(i\), the RLNC decoder performs a progressive decoding on received coded packets and de-interleaves the decoded symbols to recover the original source packets as shown in Fig.\ref{fig:RLNC-decoder}.
The RLNC decoder can partially decode an original packet belonging to a generation before all coded packets of that generation \(g\) arrive. Therefore, a couple of decoding techniques are incorporated in the proposed RLNC decoder design, such as the LU decomposition algorithm, inversion of the encoding matrix, and Gauss-Jordan elimination \cite{jones_binary_2015}. These techniques are featured in receiver design to iteratively decode the original packets and enhance the decoding efficiency. In addition, a sliding window approach \cite{khan_characterisation_2018} is used to track previously decoded original packets that belong to a single generation \(g\). In fact, the RLNC decoder does not require the received packets to be in order or need acknowledgment for the received coded packets. Moreover, the proposed decoding algorithm can be parallelized to optimize the performance and decrease the decoding stalls by implementing it on multi-core processors that are commercially available in current mobile devices.
\subsection{MIMO-LiDAL System} 

\begin{figure}[h]
    \centering
    \includegraphics[width=1.0\linewidth]{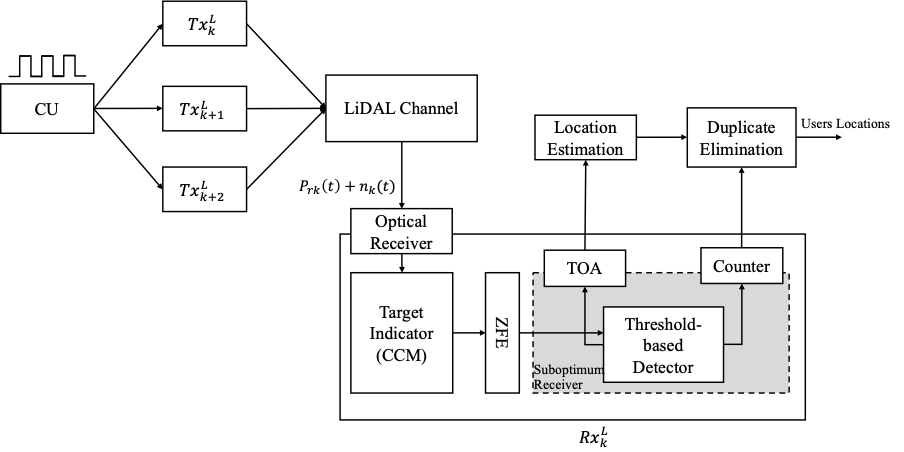}
    \caption{MIMO-LiDAL receiver design.}
    \label{fig:MIMO-LiDAL-RX-design}
\end{figure}

MIMO-LiDAL system consists of a set of LiDAL receivers that are placed on the ceiling of the room and coupled to the OWC system transmitters (i.e., eight optical APs) as shown in Fig.\ref{fig:system-model}. Each LiDAL receiver \(Rx_k^L, k=1,2,\cdots,K\), is composed of three main components: optical receiver, target indicator, and sub-optimum receiver (SOR). First, the optical receiver consists of a single photodetector, a compound parabolic concentrator (CPC), and an optical filter. Second, the LiDAL target indicator that applies the cross-correlation method (CCM) \cite{al-hameed_lidal_2019},\cite{al-hameed_artificial_2019} on the received snapshots \((S)\) of measurements to {distinguish multiple mobile users (targets) in the indoor environment observed at a given time slot \(T_s\)}. Finally, the SOR is composed of two elements: a zero-forcing equalizer (ZFE) to eliminate the effect of inter-time slots interference (ITI), and a threshold-based comparator for the target detection decision. In summary, the LiDAL receiver architecture is depicted in Fig.\ref{fig:MIMO-LiDAL-RX-design}. We assume that the MIMO-LiDAL system can utilize the same optical APs of the RLNC-NOMA communication system to generate detection and range optical pulses in time slots \(T_s\) within a pre-defined localization time window to avoid interference between detection and communication signals. Without loss of generality, we assume that an arbitrary target user \(i\) (i.e.,\(\ i\in N\)) is detected and located within the proposed MIMO-LiDAL system model at any given snapshot (i.e., time slot). { We adopt a snapshot-based model aligned with users’ nomadic behavior over LiDAL localization intervals and NOMA downlink slots. The system state is updated discretely from frame to frame to represent slow indoor mobility as a sequence of quasi-stationary snapshots.}

As a result, the communication plane is divided into spatially overlapped optical footprints \(K\) as shown in Fig.\ref{fig:system-model}. The FOV angle of the LiDAL receivers (i.e., the optical footprints \(K\)) is optimized to minimize the distance root mean square error (DRMSE) of the proposed MIMO-LiDAL system in the considered indoor environment. The minimum distinguishable distance between two arbitrary target users (localization resolution), \(\Delta R=0.3\;m\) is maintained as in \cite{al-hameed_lidal_2019}.
 
\subsubsection{LiDAL Configurations}

\begin{figure}
   \centering
    \includegraphics[width=0.75\linewidth]{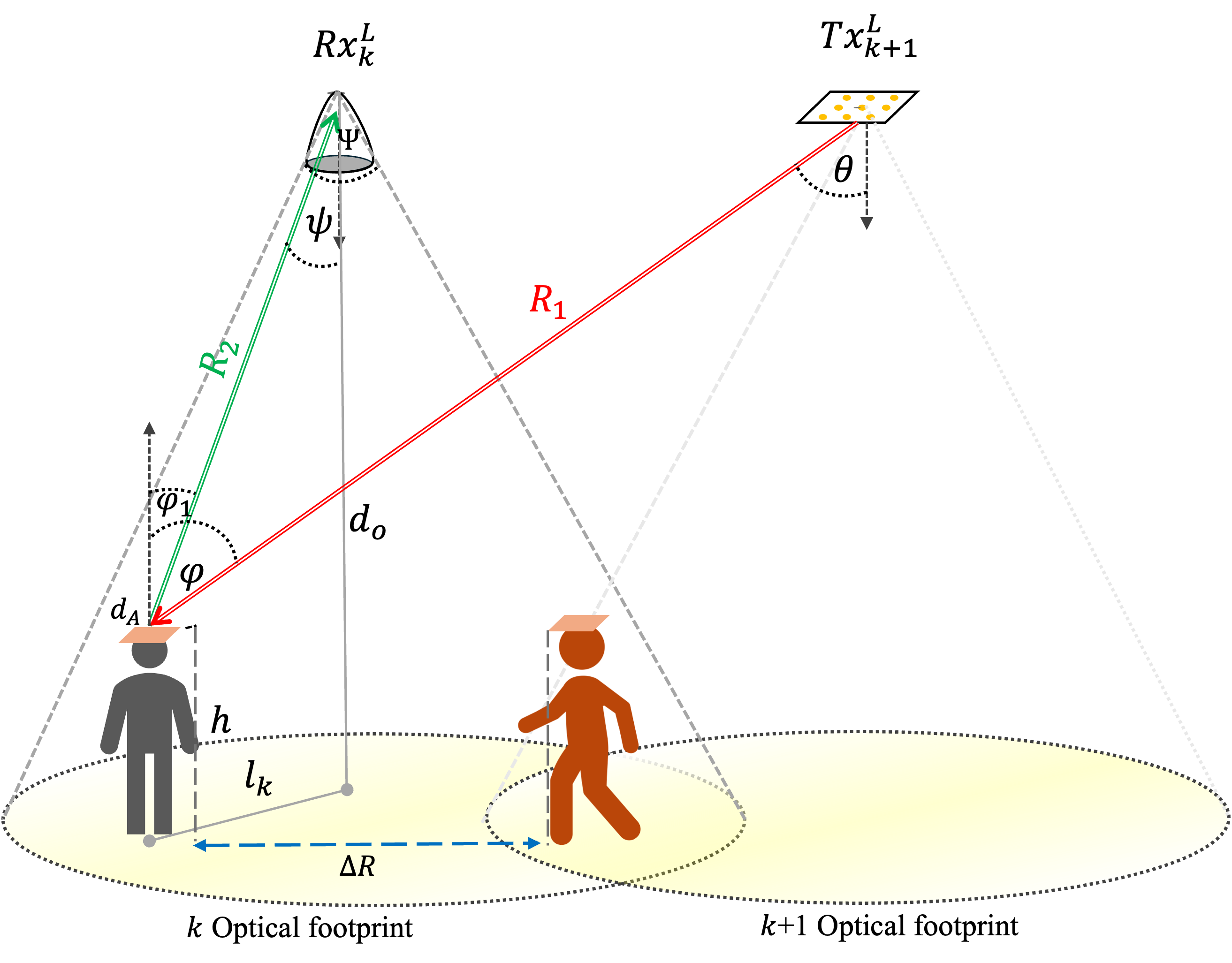}
    \caption{Bistatic LiDAL configuration.}
    \label{fig:bistatic LiDAL}
\end{figure}

\begin{figure}
     \centering
     \includegraphics[width=0.75\linewidth]{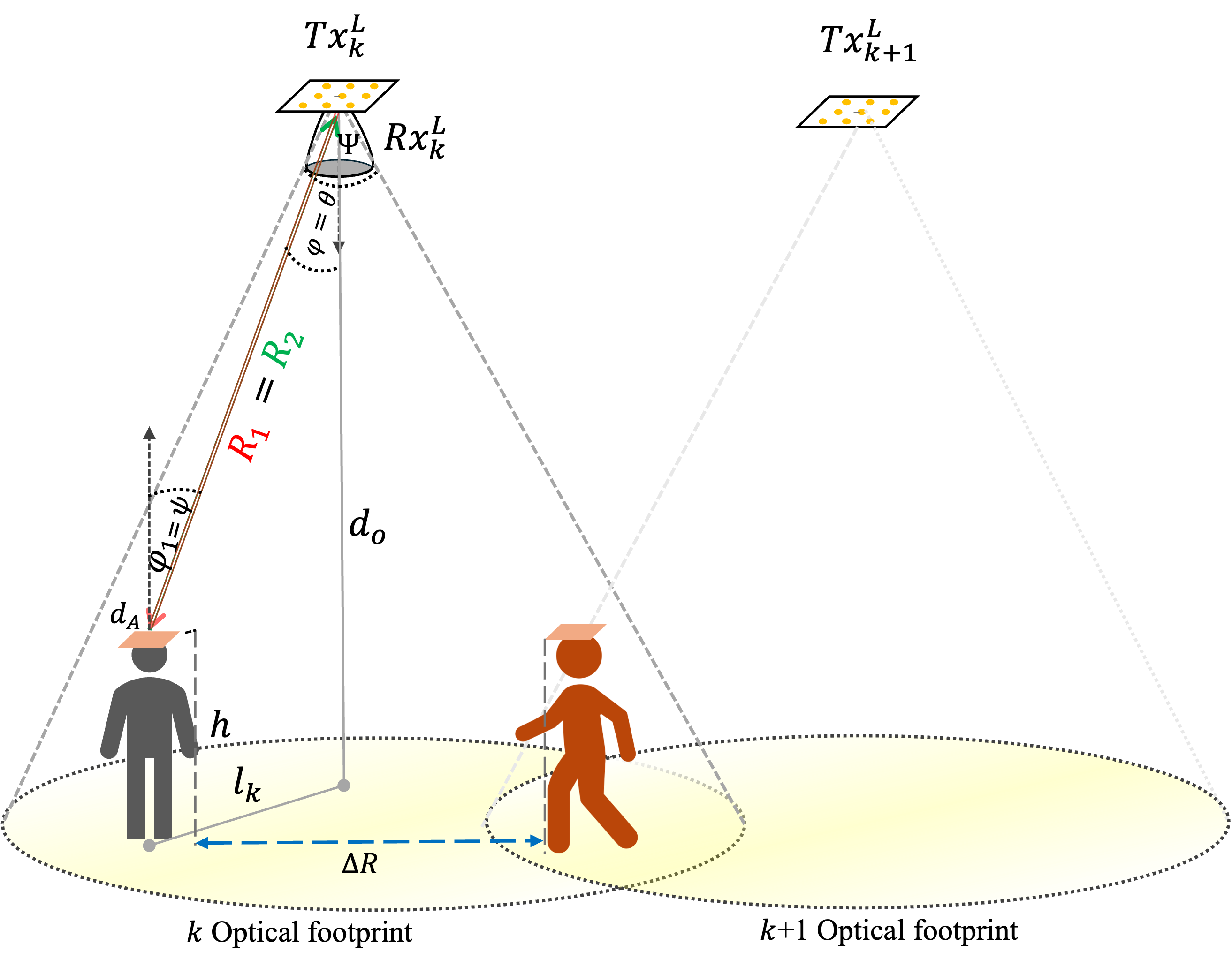}
     \caption{Monostatic LiDAL configuration.}
     \label{fig:monostatic-LiDAl}
 \end{figure}

MIMO-LiDAL system has two possible transmitter and receiver configurations to detect and locate a user inside an optical footprint. First, bistatic LiDAL which refers to a spaced transmitter and receiver configuration as illustrated in Fig.\ref{fig:bistatic LiDAL}. Second, monostatic LiDAL which refers to a collocated transmitter-receiver (i.e., transceiver) configuration see Fig.\ref{fig:monostatic-LiDAl}. In both configurations, we assume that the radiated optical signals from a transmitter and the reflected signals that obtained by a receiver follow a generalized Lambertian radiation pattern. Accordingly, the LOS optical power received from the reflected signal of a target user \(i\) by a bistatic LiDAL configuration within an optical footprint (i.e., \(k\in K\)) is given by
\begin{multline}
    Pr_{i}^{B_{LOS}}\ =\\
\begin{cases} \dfrac{(m+1)A_{PD}}{2{\pi}^2{R_1}^2{R_2}^2}{P_t d_A\rho\,  cos^m(\theta)cos(\psi)}\times\\ \quad \quad \quad \quad \quad \quad {cos(\varphi)cos(\varphi_1)T_f(\psi)g_c(\psi)}, \quad{0<\psi<\Psi^L} \\ 
 0, \quad \! \text{otherwise } \
\end{cases}\\
\label{eq:Pr-bistatic-LOS}
\end{multline}
where \(P_t\) is the pulse power of the optical signal, \(A_{PD}\) is the PD area of a LiDAL receiver \(Rx_k^L\), { \(d_A\) is the effective cross section of a target user \(i\) that denotes the circle cross-section area of the circular cylinder human body model \cite{panero_human_2014} at the assumed communication height (i.e., \(h'=0.8\) \(m\)), \(\rho\) is the reflectivity factor of the user \(i\) that represents a random variable following Gaussian distribution \(\mathcal{N}(0.72,0.3)\) according to \cite{al-hameed_lidal_2019}}. \(R_1 \) and \(R_2 \) are respectively, the LOS range distance between the transmitter \({Tx}_j^L\) and the target user \(i\), and the LOS range distance between the target user \(i\) and the receiver \({Rx}_k^L\) and \(k\neq j\). \(\theta\) and \(\varphi_1\) devote respectively, the irradiance angle of the radiated signal of the transmitter \({Tx}_j^L\) and the irradiance angle of the reflected signal of the target user \(i\). \(\varphi\) and \(\psi\) are, respectively, the incidence angle of the optical signal transmitted to the target user \(i\), and the incidence angle of the reflected signal of the target user \(i\) to the receiver \({Rx}_k^L\). \(T_f\left(\psi\right)\) and \(g_c\left(\psi\right)\) denote the optical filter and the optical concentrator gains, respectively. \( \Psi^L\) is FOV angle of the LiDAL receive \(Rx_k^L
\), and \(m\) denotes the Lambertian emission index.
On the other hand, the optical power of the reflected signal received from a target user \(i\) in a monostatic LOS configuration is given by
\begin{multline}
    Pr_{i}^{M_{LOS}} = \\
\begin{cases}\frac{(m+1)A_{PD}}{2{\pi}^2{R}^4}{P_t d_A\rho\,cos^{m+3}(\psi)T_f(\psi)g_c(\psi)},
 & {0<\psi<\Psi^L} \\
 0, & \text{otherwise}
 \end{cases}
 \label{eq:Pr-monostatic-LOS}
\end{multline}
 where \(R\) (i.e., \(R_1=R_2\) ) is the LOS range distance between the transceiver (\({Tx}_k^L\)-\({Rx}_k^L\) ) and the target user \(i\) within the optical footprint \(k\). Note that, \(\psi\) represents both the irradiance and incidence angles where (\(\psi=\theta\) and \(\psi=\varphi_1\)) of the transmitted optical signal and the reflected optical signal of the target user \(i\) in the monostatic LiDAL configuration as shown in Fig.\ref{fig:monostatic-LiDAl}.

\subsubsection{Localization method}
Time-of-arrival (TOA) technique is proposed in the MIMO-LiDAL system to estimate the location of users in the considered indoor environment. For an accurate localization of an arbitrary user \(i\) inside an optical footprint \(k\), the receiver \({Rx}_k^L\) should obtain at least three range signals that are transmitted from three LiDAL transmitters \({Tx}_j^L\). In other words, for an arbitrary user \(i\) should be in the coverage of one monostatic LiDAL subsystem (i.e., optical footprint \(k\)) and simultaneously ranged by at least two neighboring bistatic LiDAL subsystems (\(\forall j \in K^B\),\( 2\leq|{K^B}|,K^B \subset K\) and \(k\neq j\)) as shown in Fig.\ref{fig:MIMO-LiDAL-LOC}.

The time of arrival of a reflected pulse signal \(s(t)\) from target user \(i\) on a LiDAL receiver \(Rx_k^L\) within a time slot \(T_s\) can be written as
\begin{dmath}
    \mathcal{T}_i^k = \text{arg}\max_{\tau}\bigg({\int_{-T_s}^{T_s} I_i^s P_{ri}(t-T_s+\tau)\,s(t)\:\mathrm{d}t}\bigg),
\end{dmath}
where \(P_{ri}\) is the total received power of the reflected pulse signal, \(\tau\) is the time delay, and \(I_i^s\) is the CCM indicator of the presence of user \(i\) that occurred in a time interval \(T_s\) for consecutive measurements snapshots \(S\). Thus, \(I_i^s\)can be defined as
\begin{dmath} 
    I_i^s =
    \begin{cases}
       0,  &  \quad \text{if user}\; i \text{ is absent} \\
       1.  &  \quad \text{if user}\; i \text{ is present} 
    \end{cases}
\end{dmath}
The range of target user \(i\) lies inside a monostatic LiDAL (i.e., optical footprint \(k\)) is given by.
\begin{dmath}
    R_i^k =\frac{{\mathcal{T}_i^{k,j}}c}{2}\,,  \quad {k \in K, k=j}
\end{dmath}
where \(\mathcal{T}_i^{k,j}\) is the estimated TOA of the transmitter \(Tx_j^L\) signal and its reflection to a LiDAL receiver \(Rx_k^L\), and \(c\) denotes the speed of light propagation. 
Therefore, the bistatic LiDAL range between transmitter \(Tx_j^L\) and user \(i\) can be derived as
\begin{dmath}
    R_i^j ={ \mathcal{T}_i^{k,j}c - R_i^k}.  \quad { k,j \in K, k\neq j}
\end{dmath}
Let us assume the location of target user \(i\) on the communication plane is defined by coordinates vector \( X_i=\left[\begin{matrix}x_i\\y_i\\\end{matrix}\right]\)and it can be computed by the following range function.
\begin{dmath}
R_i^k = \sqrt{(x_k-x_i)^2 + (y_k-y_i)^2}, \quad {k \in K}
\label{eq:ranging-LiDAL}
\end{dmath}
where \(\left[x_k,\ y_k\right]^T\) is the location coordinates vector of the LiDAL transmitter \({Tx}_k^L\) on the ceiling plane. 
By applying the least squares method, the location vector \(X_i\) of target user \(i\) can be estimated as an intersection of minimum three circles (i.e., optical footprints \(\left|K\right|\geq3\)), and it can be expressed in matrix notation as
\begin{align}
\begin{bmatrix}\hat{x}_i\\\hat{y}_i\\\end{bmatrix} = {[A^T A]}^{-1}{A^T}B,
\label{eq:TOA-Location-estimation-fn}
\end{align}
where
\begin{dmath}
     A =\begin{bmatrix}
        {x_k - x_1} & {y_k - y_1}\\
        \vdots & \vdots\\
        {x_K - x_1} & {y_K - y_1}
    \end{bmatrix},
    \end{dmath}
    and 
    \begin{dmath}
        B = \frac{1}{2}\begin{bmatrix}
            {R_1^2- R_k^2-x_1^2+x_k^2-y_1^2+y_k^2} \\
            \vdots \\
            {R_1^2- R_K^2-x_1^2+x_K^2-y_1^2+y_K^2}
        \end{bmatrix},
    \end{dmath}
    \begin{align*}
        \text{for}\; k=2,\cdots,K.
    \end{align*}

\section{System Design Analysis}
In this section, we discuss the integrated design functions of the RLNC-NOMA and MIMO-LiDAL systems, focusing primarily on the decoding success and the localization probability, respectively. Furthermore, an unbiased location error estimator is derived for the proposed MIMO-LiDAL system. A method to group users effectively using LiDAL-assisted technique is also proposed and analyzed.
\subsection{RLNC-NOMA Decoding Success Probability}

The total success probability of the proposed RLNC-NOMA and conventional NOMA schemes is derived to be evaluated for the system model considered. Generally, the total success probability is defined as the ratio of users who correctly recover all packets in a limited number of transmissions.

Initially, by applying the RLNC-NOMA transmission methodology, the achievable data rate \(\gamma_{k,i}\) for user \(i\) located in a group \(k\) is derived as follows \cite{park_random_2015}.
\begin{multline}
    \gamma_{k,i} = \\
    \begin{cases}
          \resizebox{.95\hsize}{!}{$B_w \log_2(1+\dfrac{\mu_i \mathfrak{R}_i^2 \hbar_i^2 \mathcal{P}_t^2\alpha_i^2 }{\mathfrak{R}_i^2 h_i^2\mathcal{P}_t^2\sum^{M_k}_{j=i+1}\alpha_j^2 + \sigma_i^2}) \leq \zeta,\; 1 \leq i < M_{k} 
          $}\\
       B_w \log_2(1+\dfrac{\mu_i \mathfrak{R}_i^2  \hbar_i^2 \mathcal{P}_t^2\alpha_i^2}{\sigma_i^2}) \leq \zeta,\; i= M_k
    \end{cases}\\
      \label{eq:RLNC-NOMA-rate}
\end{multline}
where \(B_w\) denotes the PD receiver bandwidth,  \(\zeta\) is {the minimum conventional NOMA throughput to capture a transmitted packet}, \(\mathfrak{R}_k\) is the receiver PD responsivity of user \(i\) , { \(\mu_i\) is a constant to maintain the decoding proportionality under the constraints of intensity modulation and direct detection.}

We assume that the total number of successive transmissions is denoted by \(V\) . Then, to derive the total probability of success, the probability of failure to transmit an arbitrary packet to a generic user \(i\) is denoted by \(\delta_{i,v}\) and given as \cite{park_random_2015}.
\begin{align}
    \delta_{i,v} = 1 - \exp(-\varepsilon_i) {\sum_{v=1}^V \dfrac{\varepsilon_i^{v-1}}{(v-1)!}}, \quad
    \forall i \in M_k 
\end{align}
where \(v\) is the transmission index at which the successive transmission occurred.

Suppose that \(\varepsilon_{k,i}\) denotes the ratio of the optical channel gain to the power allocated to an arbitrary user \(i\) in a group \(k\) and be derived recursively as follows.
\begin{multline}
    \varepsilon_{k,i} =
    \begin{cases}
        \dfrac{\Omega_{k,i}}{\mathcal{P}_t^2\bigg(\alpha^2_i -{ \alpha^2_{i+1}\big(2^{\frac{\zeta}{B_w}}-1\big)}\bigg)},\; 1 \leq i< M_k \\
        \dfrac{\Omega_{k,i}}{\mathcal{P}_t^2\alpha_i^2 }, \quad  i= M_k
    \end{cases} \\
\end{multline}
where \(\Omega_{k,i}\) is the optical channel gain of user \(i\) in group \(k\) that can easily be derived from (\ref{eq:RLNC-NOMA-rate}). Note that \(\alpha^2_i > { \alpha^2_{i+1}\big(2^{\frac{\zeta}{B_w}}-1\big)} \) is a necessary condition to satisfy the NOMA decoding order principle.

To compute the success probability for each user \(i\) within an arbitrary group \(k\), it is first assumed that the optical AP transmits each packet twice in the conventional NOMA scenario and four coded packets in the RLNC-based NOMA to maintain fairness. Furthermore, the binomial probability density function should satisfy the requirement that each user receives at least two packets in both scenarios. Consequently, the total success probabilities for the RLNC-NOMA \(\hat{p_s}\) and the NOMA \(p_s\) schemes in the \(v\)-th successive transmission are derived from \cite{park_random_2015}, \cite{tsimbalo_reliability_2018} and given respectively by 

\begin{align}
   \hat{p_s} = \prod_{i=1}^{M_k} 1 - \bigg(\displaystyle\sum_{\lambda = 0}^{f-1}\binom{\hat{\tau}}{\lambda}(\delta_{i,v})^{\hat{\tau}-\lambda}(1-\delta_{i,v})^\lambda\bigg),
    \label{eq:RLNC-NOMA-Prob}
\end{align}
\begin{align}
    p_s = \prod_{i=1}^{M_k}(1-\delta_{i,v})^f,
    \label{eq:NOMA-Prob}
\end{align}
where \(f\) is the total number of packets in the frame (generation) and \(\hat{\tau}\) donates the total number of transmission attempts to all users of group \(k\) considered in the RLNC-based NOMA scenario. It is worth pointing out that \(v=1\) in (\ref{eq:RLNC-NOMA-Prob}) since users are assumed to receive a linearly independent coded packet at any given time.

\subsection{MIMO-LiDAL Localization Probability}
In the proposed MIMO-LiDAL system, three unique LiDAL scans are required from adjacent transceivers to detect the target user \(i\) and estimate its location. Implicitly, the probability of localizing users in the MIMO-LiDAL system is based on the independent detection probability of the monostatic LiDAL subsystem and the probability of the collaborating \(k\) bistatic LiDAL subsystems. Therefore, the probability of localizing an arbitrary target user \(i\) within overlapping optical footprints \(K\) can be derived as
\begin{align}
P_i^L = P_D^M\prod_{k=2}^{K}{P_{D}^B(k)}, \quad {i \in M_k},{k\in K}
\end{align}
where \(P_D^M\) is monostatic LiDAL detection probability of target user \(i\) and \(P_D^B\) is the detection probability of all adjacent \(k\)-bistatic LiDAL subsystems of the same target user.
The monostatic LiDAL subsystem detection probability can be derived as \cite{al-hameed_lidal_2019}
\begin{align}
P_D^M= \frac{1}{2} Q\bigg(\frac{D_{th}^M-\mu_M}{\sqrt{2}\sigma_M}\bigg).
\end{align}
The detection probability of an adjacent \(k\)-th bistatic LiDAL subsystem is given by \cite{al-hameed_lidal_2019} 
\begin{align}
    P_D^B(k) = \frac{1}{2} Q\bigg(\frac{D_{th}^B(k)-\mu_B(k)}{\sqrt{2}\;\sigma_B(k)}\bigg),
\end{align}
where \(Q (.)\) is the error function, \(\mu_M\) and \(\sigma_M\) respectively are the mean and variance of the received power in monostatic LiDAL subsystem. \(\mu_B\) and \(\sigma_B\) denote respectively, the mean and variance of the power received from the \(k\)-th bistatic subsystem. \(D_{th}^M\) and \(D_{th}^B\) denote respectively, the detection thresholds of monostatic and \(k\)-bistatic LiDAL subsystems. The detection threshold of both subsystems can be computed by \cite{al-hameed_lidal_2019}
\begin{multline}
    D_{th} = \\
   \textstyle \sqrt{\bigg({\frac{\mu^2}{(\beta_\sigma -1)^2}} + {\frac{\mu^2}{\beta_\sigma -1}} + {\frac{2\sigma^2}{\beta_\sigma -1}}\big(ln\frac{\gamma_{FP}}{\gamma_{FA}} - ln \frac{\hat{\sigma}_t}{\sigma}\big)\bigg) - \frac{\mu}{\beta_\sigma -1} }.
\end{multline}
Thus, \(\mu\) and \(\sigma\), respectively, are the mean and standard deviation of the received reflected signals in each respective LiDAL subsystem. Note that,
\begin{align}
\sigma^2 = \hat{\sigma_t}^2 + \hat{\sigma_s}^2,
\end{align}
where \(\hat{\sigma_s}\) standard deviation of the received reflected signals in noise.  \(\hat{\sigma_t}=\sqrt{\left(\ \sigma_{thr}^2+\ \sigma_{shot}^2\right)}\) denotes total of standard deviation of both thermal noise and shot noise components of an arbitrary LiDAL receiver \cite{al-hameed_lidal_2019}. \(\beta_\sigma\geq 1 \) represents the factor that measures the variance of the reflected signals due to the target coating effect (i.e., worn cloth materials and colors). This \(\beta_\sigma\) factor can be given by \cite{al-hameed_lidal_2019}
\begin{align}
    \beta_\sigma = \frac{\hat{\sigma}_s^2 + \hat{\sigma}_t^2}{\hat{\sigma}_t^2}.
\end{align}
Moreover, \(\frac{\gamma_{FP}}{\gamma_{FA}}\) is the LiDAL likelihood test-ratio threshold for SOR detection decision. knowing that \(\gamma_{FP}\) is the cost factor of missing a user presence, and \(\gamma_{FA}\) is the cost factor of falsely identifying the presence of a user in an indoor environment.

In our proposed MIMO-LiDAL system, we assume the cost of missing a user is high, then the test-ratio threshold should be set very low. Consequently, the LiDAL test-ratio threshold cost factors are set as \(\gamma_{FP}=1\) and \(\gamma_{FA}=100\) to minimize the probability of error in detecting and estimating the location of the user in the considered indoor environment \cite{al-hameed_lidal_2019}. 
\begin{figure}
     \centering
     \includegraphics[width=0.75\linewidth]{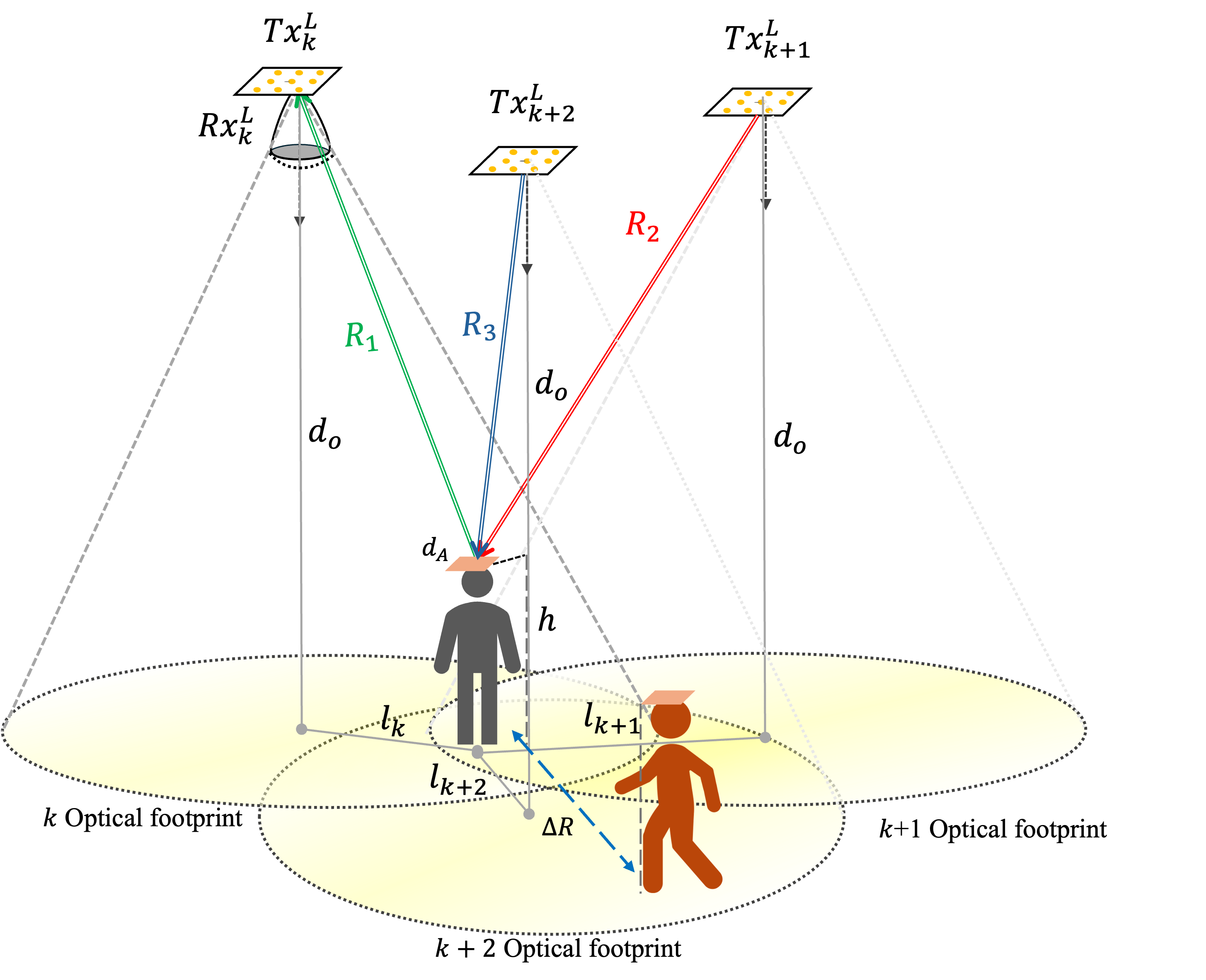}
     \caption{MIMO-LiDAL localization system.}
     \label{fig:MIMO-LiDAL-LOC}
 \end{figure}
\subsection{MIMO-LiDAL Cram\'er--Rao Lower Bound }
The minimum error of arbitrary user location estimation in the MIMO-LiDAL system model can be analytically derived by the Cram\'er-Rao Lower Bound (CRLB) \cite{godrich_target_2010}. CRLB is known as the lower bound on the mean square error of the unbiased parameter estimator. However, CRLB is widely used to theoretically evaluate the accuracy of location estimation using wireless technologies.

In this work, we derived CRLB to provide a baseline of the accuracy of the proposed localization method in the proposed LiDAL-assisted OWC system.
The total LOS optical power received from user-reflected signals by the LiDAL receiver can be expressed by 
\begin{align}
   {Pr}^{Total}= Pr_{1}^{M_{LOS}} + \sum_{k=2}^K Pr_{k}^{B_{LOS}} + \sum_{k=1}^K n_k, \quad {\forall k \in K}
   \label{eq:total-pr-LiDAL}
\end{align}
where \(Pr^{M_{LOS}}\) and \(Pr^{B_{LOS}}\) represent the LOS optical power received from monostatic and \(k\)-bistatic subsystems, respectively. {\(n_k\) denotes the diffuse and non-line-of-sight (NLOS) reflections of the overlapped optical footprints \(K\) and the AWGN of the corresponding LiDAL receiver \(Rx_k^L\) (i.e., \(k=1\)) over a time interval \(T_s\)}. Knowing that in (\ref{eq:Pr-bistatic-LOS}), \(cos(\theta_k)=\frac{(d_0-h')}{R_k^2}\) for \(k=2,\cdots,K\) and in (\ref{eq:Pr-monostatic-LOS}) \(cos(\psi_1)= \frac{(d_0-h')}{R_1^2}\) for \(k=1\). The range distance \(R_k\) in (\ref{eq:ranging-LiDAL}) can be rewritten as
\begin{align}
    R_k= \sqrt{l_k^2 + (d_0 -{h'})^2}, \quad {\forall k \in K}
\end{align}
where \(l_k\) is the distance between a projection of a transmitter \({Tx}_k^L\) on the communication plane and a target user \(i\). \(d_0\) denotes the ceiling height of the room, and \(h'\) represents the vertical height of the effective cross section of a target user \(i\), as shown in Fig.\ref{fig:MIMO-LiDAL-LOC}.
Moreover, the optical power of the LOS range signals received from the monostatic and \(k\)-bistatic subsystems can be reformulated, respectively, as 
\begin{align}
    Pr_{1}^{M_{LOS}} = \Delta_l \frac{\eta^{m+3}}{(l_1^2 + \eta^2)^{\frac{m+7}{2}}}, \quad {k=1}
\end{align}
and
\begin{align}
    Pr_{k}^{B_{LOS}} = \Delta_l \frac{\eta^{m+3}}{(l_1^2+\eta^2)^2(l_k^2 +\eta^2)^{\frac{m+3}{2}}},  \quad {k=2,\cdots,K}
\end{align}
where \(\Delta_l=P_t\frac{(m+1)A_{PD}}{2\pi^2}d_A\rho\; T_f(\psi)g_c(\psi)\).

The total optical power of the LOS signals received by the LiDAL receiver can be written as a function of the Euclidean distance \(f(l_k)\) of a transmitter \({Tx}_k^L\) and the location of a target user \(i\), and the height difference \(\eta\).
Accordingly, the first two terms in (\ref{eq:total-pr-LiDAL}) can be denoted in a vector notation as
\begin{align}
    \boldsymbol{f}(\boldsymbol{l}) =\begin{bmatrix}
        f_1(l_1) \\
        \vdots \\
        f_k(l_k)\\
        \vdots\\
        f_K(l_K)
    \end{bmatrix}, \quad {\forall k \in K}
\end{align}
and the last noise term is denoted by

\begin{align}
    \boldsymbol{n}= \left[\begin{matrix}
        n_1 \\
        \vdots \\
        n_k \\
        \vdots \\
        n_K \\
    \end{matrix}\right]. \quad {\forall k \in K}
\end{align}
Subsequently, the first derivative of \(f(l)\) is given by
\begin{align}
    \frac{\partial \boldsymbol{f}(\boldsymbol{l})}{\partial \boldsymbol{l}^T} = \begin{bmatrix}
        \frac{\partial f_1{(l_1)}}{\partial l_1} & \cdots & \frac{\partial f_1{(l_1)}}{\partial l_K}\\
        \vdots & \ddots & \vdots \\
         \frac{\partial f_K{(l_K)}}{\partial l_1} & \cdots & \frac{\partial f_K{(l_K)}}{\partial l_K}
    \end{bmatrix}.
    \quad  \quad {\forall k \in K}
\end{align}
It is worth mentioning that \(\frac{\partial {f_i}(l_i)}{\partial{l_j}} = 0\) when \(i\neq j\).
Thus,\(\frac{\partial f(l)}{\partial l^T}\) is a diagonal matrix, and it can be written as
\begin{align}
    \frac{\partial \boldsymbol{f}(\boldsymbol{l})}{\partial \boldsymbol{l}^T} = - \Delta_l\; \textit{diag} {\begin{bmatrix}
        \frac{\Pi^M l_1}{(l_1^2 + \eta^2)^{\frac{m+9}{2}}}\\
        \frac{\Pi^B l_2}{(l_1^2 + \eta^2)^2(l_2^2 + \eta^2)^{\frac{m+5}{2}}}\\
        \vdots\\
        \frac{\Pi^B l_k}{(l_1^2 + \eta^2)^2(l_k^2 + \eta^2)^{\frac{m+5}{2}}}\\
        \vdots \\
        \frac{\Pi^B l_K}{(l_1^2 + \eta^2)^2(l_K^2 + \eta^2)^{\frac{m+5}{2}}}
    \end{bmatrix}}^T,
    \quad {\forall k \in K}
    \label{eq:partial-distance-formula}
\end{align}
where \(\Delta_l\), \(\Pi^M = (m+7)\eta^{m+3}\), and \(\Pi^B = (m+3)\eta^{m+3}\) are known constants.

Let \({X}_k^{tx} = \begin{bmatrix}
    x_k \\
    y_k
\end{bmatrix}\) is the location vector of LiDAL transmitter \({Tx}_k^L\).
 The distance between the projected location of LiDAL transmitter \({X}_k^{tx}\) and the location of target user \({X}_i\) on the communication plane can be computed by
\begin{align}
\left\{
    \begin{array}{cc}
         l_1^2 = (x_i - x_1)^2 + (y_i - y_1)^2, &  \\
        \vdots & \\
        l_K^2 = (x_i - x_K)^2 + (y_i - y_K)^2.
        \end{array}
        \right. \quad {\forall k \in K}
\end{align}
Then, we have
\begin{align}
    \frac{\partial \boldsymbol{l}}{\partial \boldsymbol{X}^T} = \begin{bmatrix}
        \frac{\partial l_1}{\partial x_i} &  \frac{\partial l_1}{\partial y_i} \\
        \vdots & \vdots \\
        \frac{\partial l_K}{\partial x_i} & \frac{\partial l_K}{\partial y_i}
    \end{bmatrix} = {\begin{bmatrix}
        \frac{x_i-x_1}{l_1} & \frac{y_i-y_1}{l_1} \\
        \vdots & \vdots \\
         \frac{x_i-x_K}{l_K} & \frac{y_i-y_K}{l_K} \\
    \end{bmatrix}}. 
\end{align}
Accordingly, the CRLB of error of location estimation is given by 

\begin{align}
    B^{L}(\boldsymbol{X}) = {\begin{bmatrix}
        \bigg(\frac{\partial \boldsymbol{f}\big(\boldsymbol{l}\big)}{\partial \boldsymbol{l}^T}\bigg)^T\mathbb{Q}^{-1}\bigg(\frac{\partial \boldsymbol{f}\big(\boldsymbol{l}\big)}{\partial \boldsymbol{X}^T}\bigg)
    \end{bmatrix}}^{-1},
    \label{eq:CRLB-equation}
\end{align}
where \(\mathbb{Q} = E[\boldsymbol{n}\boldsymbol{n}^T]\) denotes the noise covariance matrix. By applying the chain rule
\(\frac{\partial f(l)}{\partial X^T} = \frac{\partial f (l)}{\partial l^T} \frac{\partial l}{\partial {X}^T}\), we obtain
\begin{align}
    \frac{\partial \boldsymbol{f}\big(\boldsymbol{l}\big)}{\partial \boldsymbol{X}^T} = - \Delta_l \begin{bmatrix}
        \frac{\Pi^M l_1(x_i-x_1)}{(l_1^2 + \eta^2)^{\frac{m+9}{2}}} & \frac{\Pi^M l_1(y_i-y_1)}{(l_1^2 + \eta^2){\frac{m+9}{2}}} \\
        \frac{\Pi^B l_2(x_i-x_2)}{(l_1^2 + \eta^2)^2(l_2^2+\eta^2)^{\frac{m+5}{2}}} & \frac{\Pi^B l_2(y_i-y_2)}{(l_1^2 + \eta^2)^2(l_2^2+\eta^2)^{\frac{m+5}{2}}}\\
        \vdots & \vdots \\
        \frac{\Pi^B l_k(x_i-x_k)}{(l_1^2 + \eta^2)^2(l_k^2+\eta^2)^{\frac{m+5}{2}}} & \frac{\Pi^B l_k(y_i-y_k)}{(l_1^2 + \eta^2)^2(l_k^2+\eta^2)^{\frac{m+5}{2}}} \\
        \vdots & \vdots \\
        \frac{\Pi^B l_K(x_i-x_K)}{(l_1^2 + \eta^2)^2(l_K^2+\eta^2)^{\frac{m+5}{2}}} & \frac{\Pi^B l_K(y_i-y_K)}{(l_1^2 + \eta^2)^2(l_K^2+\eta^2)^{\frac{m+5}{2}}} \\
    \end{bmatrix}.
    \label{eq:partial-location-distance}
\end{align}
The CRLB of location estimation error in our proposed MIMO-LiDAL system can be obtained by substituting both equations (\ref{eq:partial-distance-formula}) and (\ref{eq:partial-location-distance}) in \({B^L(\boldsymbol{X})} \). In particular, CRLB estimated parameter is utilized to enable the power allocation of the integrated RLNC-NOMA communication system that serves multiple users within the realistic indoor environment considered.

\subsection{LiDAL-assisted User Association Scheme}
To increase the throughput of users with poor channel gains (low location orders), it is beneficial to pair them with users with high channel gains (high location orders). The reason is that users with high channel gains can still achieve a higher data rate even if they have been assigned to lower power levels while making a large portion of the optical power of AP \(k\) available to users with weak channel gains. 

Consequently, we propose a LiDAL-based suboptimal user association algorithm that consists of three main operations: determine the adjacent APs (\(\forall k \in K^* , K^* \subseteq K \)) that have the minimum Euclidean distance \(f_i(k)\) to the estimated locations of OWC users on the communication plane. Then, sorting user locations according to a CRLB location error estimate \(B^L_{k}(X_i)\) and dividing them into two classes: Class I consists of users with low estimated location errors, and class II consists of users with high estimated location errors. Next, sort these users according to their access distance \(\Lambda^{k}_i\) between their locations (estimated locations) and the potential adjacent APs. Finally, apply the D-NLUPA method \cite{islam_resource_2018} to pair users from both classes based on their largest difference in access distances \(\Lambda_i^k\), to be assigned to a single adjacent optical AP (i.e.,\(k^* \in K^*\)) with the lowest estimated location error for this group. In summary, the proposed user association algorithm is  given in Algorithm \ref{alg:LiDAl-NOMA-cls}.

\begin{algorithm}[ht]
\caption{LiDAL-assisted RLNC-NOMA User association }\label{alg:LiDAl-NOMA-cls}
\begin{algorithmic}[1]
\State Initialize {$ N \gets \text{OWC users}$}
\State Initialize {$ K \gets \text{ Optical APs}$}
\For{$i \gets 1, N$}
\State Compute the Euclidean distance \(f_i(k) ,\forall k \in K\) of the location of user \(i\)  on the communication plane to all optical APs. \
\State Find the feasible sets of adjacent APs \(K^* \subset K\) that have the minimum Euclidean distance \(f_i(k)\) to user \(i\).
\EndFor
\For{$k \gets 1, K^*$}
\State Sort all the potential users \(M^k\) of optical AP \(k\) according to their location CRLB error estimation \(B^L_{k}(X_m)\; \forall m \in M^k\).
\State Divide the  \(M^k\) users into two classes (i.e., Class I and Class II).
\For{$m \gets 1, M^k_j, j=\{1,2\}$}
\State Sort each class \(j\) users \(M^k_j\) according to their access distance to optical AP \(k\).
 \begin{align*}
        \Lambda^k_1 \leq \Lambda^k_2 \leq \cdots \leq \Lambda^k_m \leq \Lambda^k_{m+1} \leq ... \leq \Lambda^k_{M_j}
    \end{align*}
\EndFor
\State Pair the all \(M^k\) users in both classes according to D-NLUPA pairing method \cite{islam_resource_2018}.
\EndFor
\State Assign the group of paired users to the AP \(k^* \in K^*\) that have the minimum average estimated location errors of the users in this group.
\State Ensure no duplicate users in the group assigned to the optical AP \(k^*\).
\end{algorithmic}
\end{algorithm}
\section{Simulation and Performance Evaluation}
\subsection{Simulation Setup}
In Fig.\ref{fig:system-model}, eight RGB-LD APs are used to serve communication users and estimate their locations within the formed overlapped optical footprints that meet the required illumination and eye safety standards (i.e., \(K=8\)). From equation (88) in  \cite{al-hameed_lidal_2019}, we derived the maximum number of users that can be detected and localized using the configuration of the MIMO-LiDAL system in Table \ref{tab:LiDAL-NOMA-system-setup} (i.e., \(N=32\)). We consider these users to be uniformly and randomly distributed on the communication plane. Consequently, each set of four users would initially be assumed to be located within each optical footprint.
We considered a realistic indoor environment with reflections generated by user skin color, cloth materials (i.e., cotton, polyester), indoor surfaces and objects (i.e., walls, furniture, door, and ceiling). 

The reflections of these elements can be represented as a combination of diffuse and specular reflectors. In this work, a stochastic process (i.e., normalized Gaussian distribution) of diffuse reflectivity is assumed and referred to by the factor \(\rho\) in equations (\ref{eq:Pr-bistatic-LOS}) and (\ref{eq:Pr-monostatic-LOS}) to model the reflections of target users, objects, and surfaces that dominate the reflected light beams in the considered indoor space \cite{al-hameed_lidal_2019}. Moreover, Lambertian first- and second-order diffuse light reflections are collected and the impulse response of NLOS optical pluses is traced from the effective cross section area \(d_A\) of an arbitrary target user, object, or surface within the proposed indoor environment. A 300 MHz optical receiver bandwidth and zero forcing equalizer (ZFE) are utilized in the LiDAL receiver design to meet the proposed minimum LiDAL resolution \(\Delta R = 0.3\) \(m\). Therefore, 378 locations on the communication plane are considered to represent all possible user locations that the MIMO-LiDAL system can detect and without detection and/or counting ambiguity due to ITI. Nomad mobility is assumed (i.e.,  \(\hat{v} \in [0.5,2]\;m/s\) ) where each user can randomly change his location at most six neighbour locations (i.e., \(\hat{v}/\Delta R\)) in a second. Note that the same receiver responsivity factor \(\mathfrak{R}\) is assumed for both user receivers and mounted LiDAL system receivers in terms of thermal and shot noises. Additionally, we assume an identical vertical height difference \(\eta\) for both the PD of the user communication receiver \(A_{PD}\) and the user detection cross section \(d_A\) for localization ( refer to equations (\ref{eq:Pr-monostatic-LOS}) and (\ref{eq:Pr-bistatic-LOS})). 

During communication, each generation of packets is encoded in GF\((2^8)\), modulated with NRZ-OOK, and superimposed with other generation symbols transmitted by an AP \(k\). For SIC, the superimposed received signals are estimated and detected using maximum-likelihood detection. The sparse RLNC scheme \cite{pedersen_kodo_2011} generates fixed-length coded packets (\(f=3\)) with random coding vectors to decode over the estimated erasure channel. The remaining RLNC-NOMA and MIMO-LiDAL parameters are given in Table \ref{tab:LiDAL-NOMA-system-setup}.
\begin{table}[h!tbp]
\caption{System setup parameters}
\centering
\begin{tabular}{|>
{\raggedright\arraybackslash}p{0.5\linewidth}|>{\centering\arraybackslash}p{0.4\linewidth}|}
\hline
\textbf{Parameters} & \textbf{Configurations } \\
\hline
Room dimensions (\(x \times y \times z\))   & \(4 \times 8 \times 3\) \(m^3\) \cite{al-hameed_artificial_2019}\\
\hline
Finished-wood,walls, floor and ceiling reflectivity factors, & 0.55 ,0.8, 0.3, 0.8  \cite{al-hameed_artificial_2019}\\
\hline
Effective user cross section, \(d_A\) & \(0.07\) \(m^2\)\\
\hline

Total transmitted optical power, \(P_t\)    & 18 W \cite{al-hameed_artificial_2019} \\
\hline
Transmitter semi-angle at half power beamwidth, \(\Phi_{1/2}\)    & 75\textsuperscript{o} \cite{al-hameed_artificial_2019}\\
\hline
 APs locations (x,y,z) & (1,1,3),(1,3,3),(1,5,3),(1,7,3), (3,1,3),(3,3,3),(3,5,3),(3,7,3) \cite{al-hameed_artificial_2019}\\
\hline
Azimuth  & 0\textsuperscript{o} \\
\hline
Elevation & 90\textsuperscript{o} \\
\hline
Optical receiver PD area, \(A_{PD}\)    & 20 mm\textsuperscript{2} \cite{al-hameed_artificial_2019} \\
\hline
PD filter gain, \(T_f(\psi)\)    & 1 \cite{al-hameed_artificial_2019} \\
\hline
Optical concentrator gain, \(g_c(\psi)\)    & 3.77 \\
\hline
PD responsivity, \(\mathfrak{R}\)    & 0.4 A/W \cite{al-hameed_artificial_2019} \\
\hline
Concentrator reflective index, \(n\)   & 1.7 \cite{al-hameed_artificial_2019}\\
\hline
\multicolumn{2}{|c|}{\textbf{MIMO-LiDAL system}} \\
\hline
LiDAL receiver FOV,\(\Psi^L\)   & 54\textsuperscript{o} \\
\hline
Transmitted pulse width    & 2 ns \cite{al-hameed_artificial_2019} \\
\hline
LiDAL receiver time slot    & 2 ns \cite{al-hameed_artificial_2019} \\
\hline
LiDAL receiver bandwidth,\(B_w^L\)  & 300 MHz \cite{al-hameed_artificial_2019} \\
\hline
The optical footprint diameter  & 2.6 m \\
\hline
The maximum optical footprints overlap  & 0.6 m \\
\hline
Time bin duration  &   0.01 ns \cite{al-hameed_artificial_2019}\\
\hline
\multicolumn{2}{|c|}{\textbf{RLNC-NOMA system }} \\
\hline
User cross section height, \(h'\)  & 0.8 m \\
\hline
User receiver FOV,\(\Psi_c\)    & 40\textsuperscript{o} \cite{yin_performance_2015} \\
\hline
Elevation & 90\textsuperscript{0} \\
\hline
Azimuth & 0\textsuperscript{0} \\
\hline
User receiver bandwidth,\(B_w\)  & 20 MHz \cite{yin_performance_2015} \\
\hline
RLNC generation size, \(f\) & 3 packets \cite{park_random_2015} \\
\hline
Minimum NOMA user throughput,\(\zeta\) & 0.5 bits/s.\({Hz}^{-1}\) \cite{yin_performance_2015}\\
\hline
\end{tabular}
\label{tab:LiDAL-NOMA-system-setup}
\end{table}

\begin{figure}[h!tbp]
    \centering
    \includegraphics[width=1.0\linewidth]{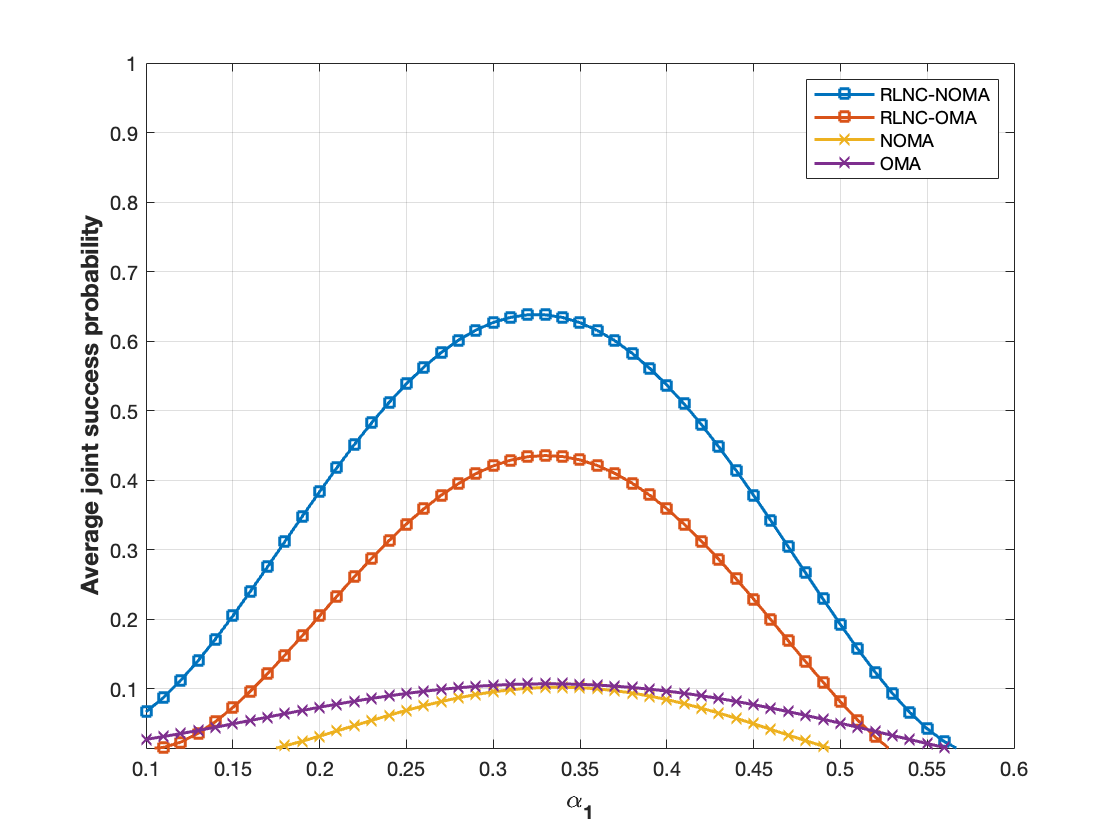}
    \caption{RLNC-NOMA and RLNC-OMA packet success probability versus the power allocation coefficient \(\alpha\).}
    \label{fig:RLNC-NOMA-OMA-succss}
\end{figure}
\subsection{RLNC-NOMA Decoding Success Probability}

{ Fig.\ref{fig:RLNC-NOMA-OMA-succss} illustrates the probability of joint success of the proposed RLNC-based NOMA and OMA schemes in relation to the power allocation coefficient \(\alpha\). This coefficient is assigned to the user with the lowest estimated location order (i.e., \(i=1\)) among a group \(k \in K\) of four sorted users according to the proposed LiDAL-NOMA association scheme (see Algorithm 1). The y-axis represents probability of all four users in a group \(k\) that successfully decode their intended packets. Noting that the four users are associated to a single optical AP \(k \in K\). The analysis of the depicted results shows that both RLNC-NOMA and RLNC-OMA have in overall the highest joint success probability compared to traditional NOMA and OMA schemes for all designated group users.}

{We found that using RLNC-based NOMA and OMA transmission schemes are more resistant in high-noise regime (i.e., \(\alpha_1 \ll 0.33\)), which is due to the recovery of the original packets are from superimposed coded packets that reduced the overall packet error rate.
Furthermore, both the performance of RLNC-based NOMA and OMA was degraded in high-interference regime but still outperforms the conventional NOMA and OMA when the assigned power allocation coefficient to the lowest location order user increases (i.e., \(\alpha_1 \gg 0.33\)). However, RLNC-based schemes provide the highest joint success probability compared to conventional ones in scenarios where power is carefully allocated and balanced among users (i.e., \(\alpha_1 = 0.33\)) of the same group under NOMA principles.}

\subsection{MIMO-LiDAL Localization Probability}

\begin{figure}
  \centering
  \subfloat[\(x\)-axis\label{1a}]{%
       \includegraphics[width=0.5\linewidth]{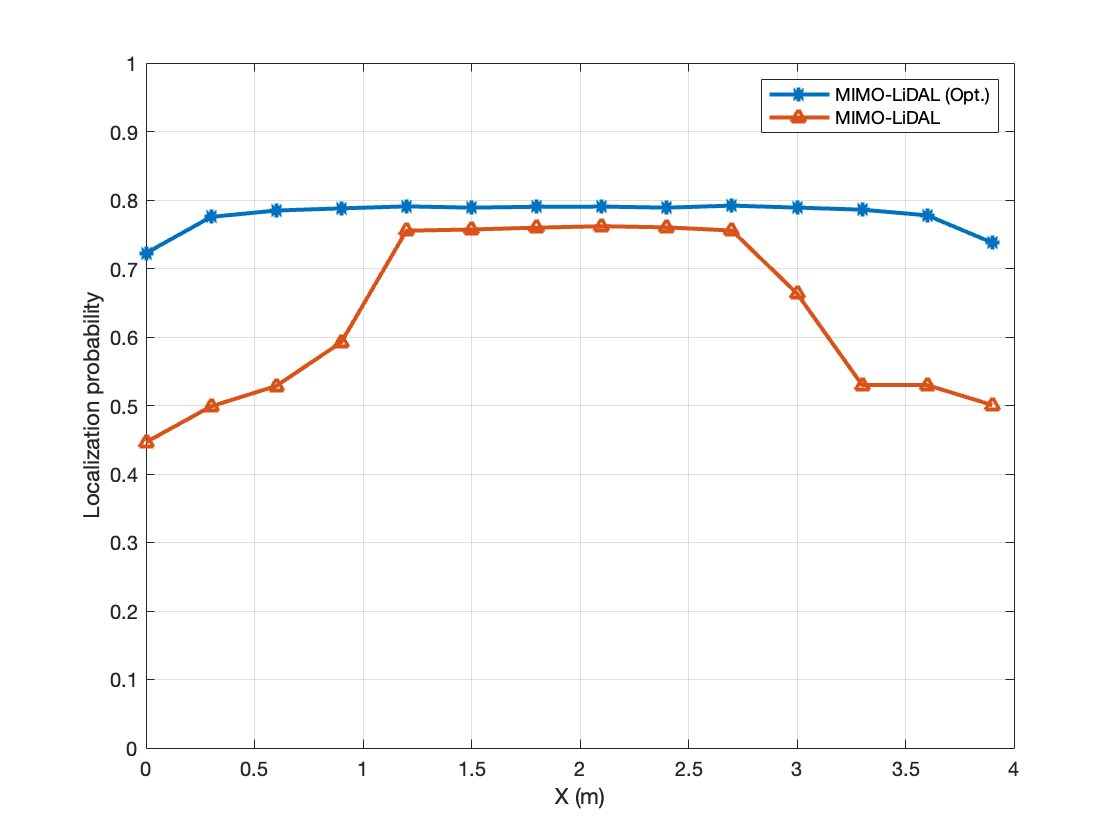}}
    \hfill
  \subfloat[\(y\)-axis\label{1b}]{%
        \includegraphics[width=0.5\linewidth]{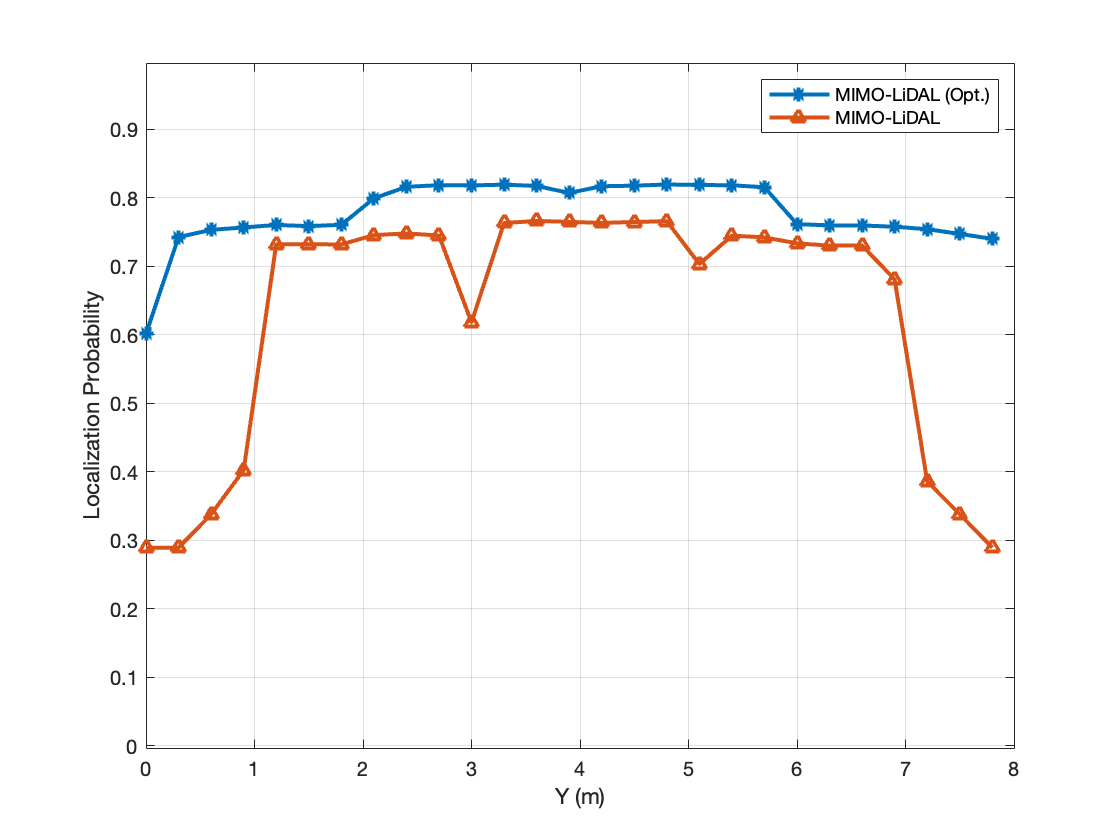}}
        \caption{PDF of MIMO-LiDAL localization on \(x\) and \(y\) -axis, respectively.}
        \label{fig:LiDAL-NOMA-Localisation-probability}
\end{figure}

{Here, we evaluate the probability of localization of our proposed MIMO-LiDAL system in the considered indoor environment and compared to the original MIMO-LiDAL system proposed in \cite{al-hameed_lidal_2019} for the same environment. In this work, we optimize two parameters of the original LiDAL receiver design, the LiDAL receiver FOV angle \(\Psi^L\), and the TOA location estimator of the MIMO-LiDAL localization system model. However, we maintain the remaining configurations of the MIMO-LiDAL system, such as the generation of optical pulses, the location of the transceiver, the optical detection parameters, and the CCM detection scheme that are specified in Table 9 \cite{al-hameed_lidal_2019}.} 
For the FOV angle of the LiDAL receiver, an exhaustive search is applied to find the optimal angle \(\Psi^L=54^o\) for the optical signals received in the formed and overlapped optical footprints while reducing the effect of the NLOS reflections of the first and second order diffuse signals generated by indoor objects and surfaces. For the TOA location estimator, the biconjugate gradient method \cite{zhou_joint_2019} is used to numerically approximate the quadratic terms in (\ref{eq:TOA-Location-estimation-fn}). The results in Fig.\ref{fig:LiDAL-NOMA-Localisation-probability} show that the probability of localization is improved on average by 23\%  by the proposed design of the MIMO-LiDAL system in both dimensions (x,y) of the indoor communication plane compared to the original MIMO-LiDAL system design.
\subsection{MIMO-LiDAL Localization CRLB and Accuracy}
\begin{figure}
    \centering
    \includegraphics[width=1.0\linewidth]{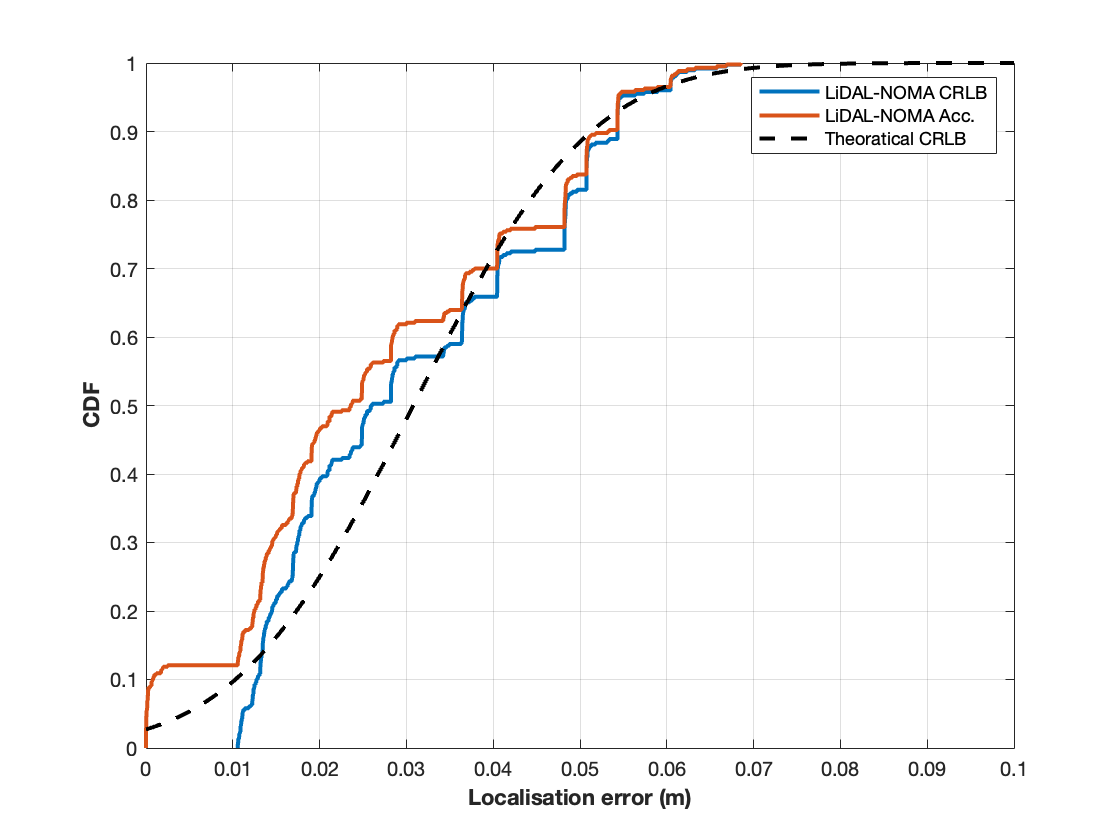}
    \caption{CDF of MIMO-LiDAL Localization CRLB vs accuracy for the proposed indoor OWC system.}
    \label{fig:CDF-CRLB-LiDAL-NOMA}
\end{figure}

We numerically evaluated the CRLB derived for the proposed MIMO-LiDAL system, as depicted in Fig.\ref{fig:CDF-CRLB-LiDAL-NOMA} for the indoor environment considered. The results highlight the distribution of the CRLB error estimations of the localization of the MIMO-LiDAL system across the communication plane while maintaining the assumption of the minimum detection resolution constraint \(\Delta R= 0.3\) \(m\). Furthermore, we computed the noise covariance matrix \((\mathbb{Q}\)) in (\ref{eq:CRLB-equation}) that consisted of ambient background noise, ambient reflections from the walls and surfaces of oblique objects (i.e., doors and tables), receiver noise components, and interfered LOS optical pulses (i.e., detection optical pulses and their reflections within overlapped bistatic subsystems). 

To validate the results of the CRLB analysis, we compared it to the localization accuracy of the MIMO-LiDAL system at all potential user locations on the communication plane, as demonstrated in Fig.\ref{fig:CDF-CRLB-LiDAL-NOMA}. The CDF graph depicts a close correlation between the CRLB bound and the localization accuracy of the proposed MIMO-LiDAL system, highlighting that the accuracy for 95\% of user locations is under 6 cm. Furthermore, the findings indicate that the CRLB function is suitable as a distance-based parameter to approximate the state of the user channel.
\subsection{LiDAL-assisted User Association}

\begin{figure}
    \centering
    \includegraphics[width=1.0\linewidth]{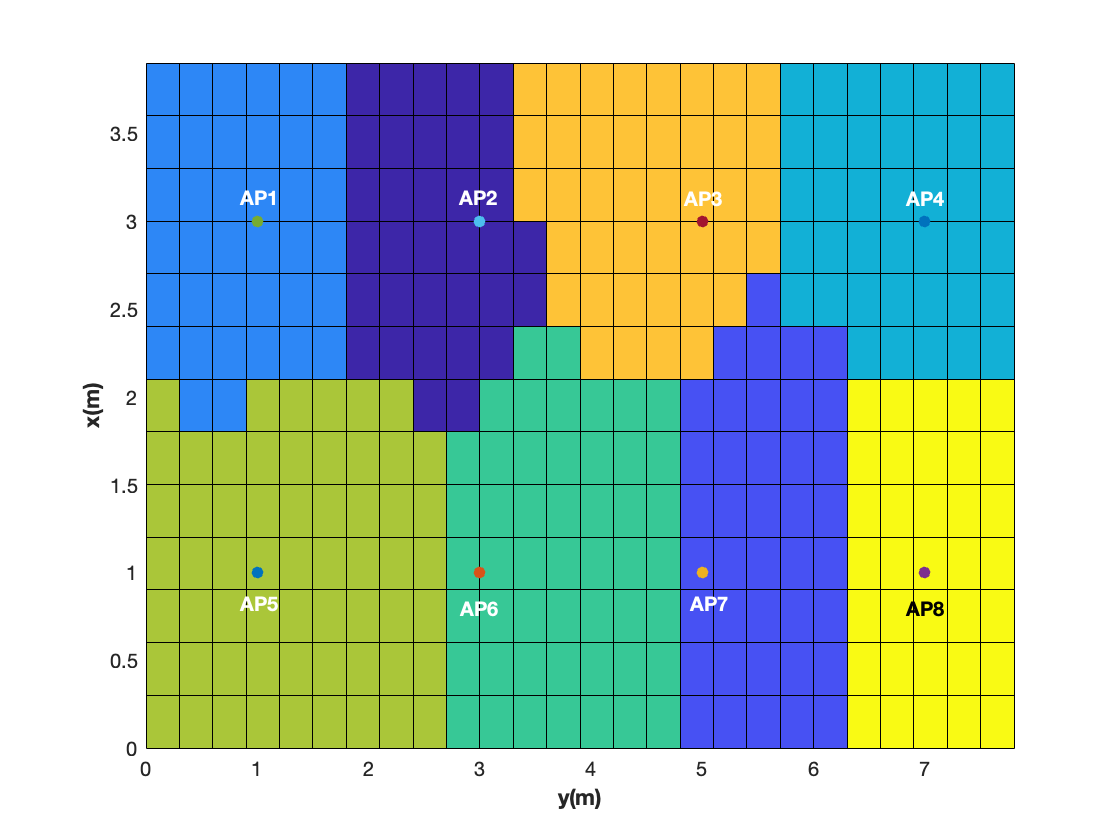}
    \caption{The associated OWC user locations to APs using the LiDAL-assisted user association scheme.}
    \label{fig:LiDAL-NOMA-Clusters}
\end{figure}

In this section, we evaluate the proposed LiDAL-assisted user association algorithm using the simulated indoor environment of the proposed RLNC-NOMA OWC system. Initially, we determined the possible locations in which an arbitrary user \(i\) can be detected and localized on the communication plane using the proposed MIMO-LiDAL system. As a result, we found that 378 possible estimated locations satisfy the minimum localization resolution condition \(\Delta R=0.3\) \(m\) on the communication plane in the proposed indoor environment. Second, we applied the user association scheme in Algorithm \ref{alg:LiDAl-NOMA-cls} on these possible locations to dynamically divide them into groups and assign each formed group to one of the system optical APs (i.e., \(K=8\)). The distribution of these locations within the generated groups ranges from 10.32\% to 15.87\%. These formed groups satisfy the three necessary conditions of the proposed algorithm: the minimum Euclidean distance between an arbitrary user \(i\) in these 378 possible locations and the optical APs of the system, the access distances (i.e., the LOS distance between the receiver of user \(i\) and the potential APs \(K^*\)) sorted in ascending order and assigned to optical AP \(k^*\) with the lowest average in estimated user location errors. Fig.\ref{fig:LiDAL-NOMA-Clusters} depicts the locations served by each optical AP \(k\) using our proposed association algorithm.
\subsection{LiDAL-assisted RLNC-NOMA sum rate}
\begin{figure}
    \centering
    \includegraphics[width=1.0\linewidth]{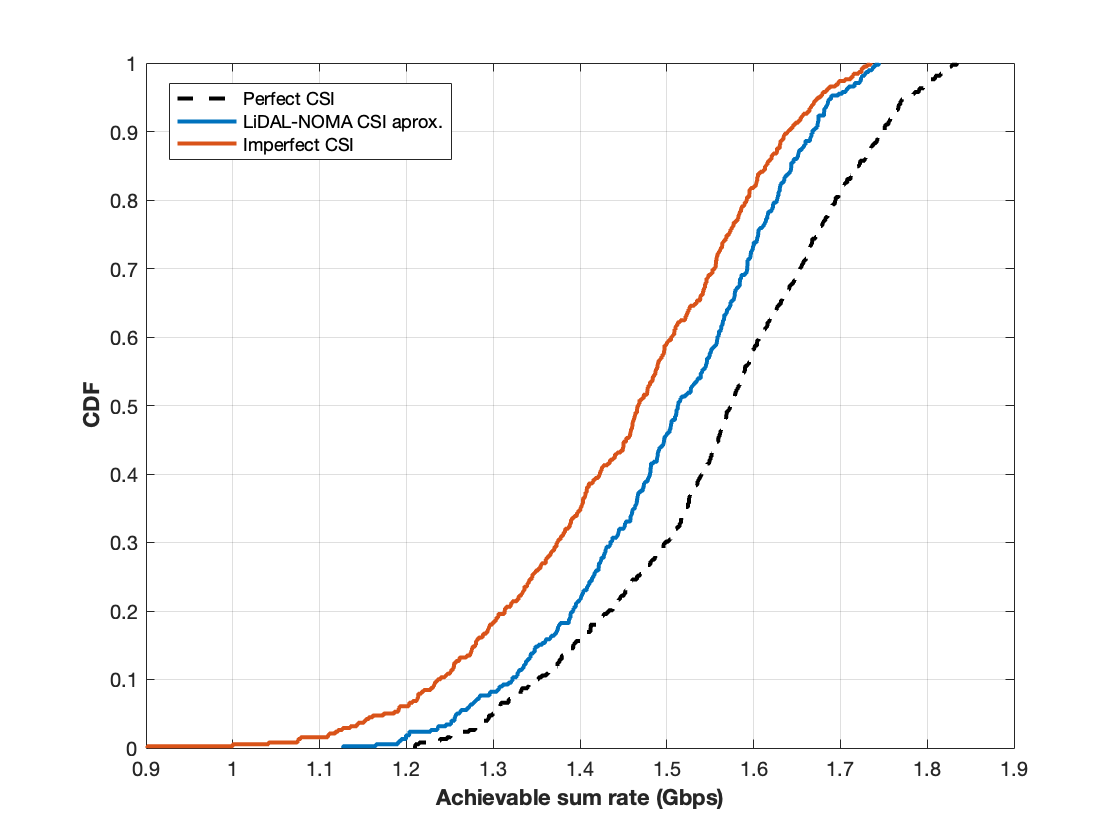}
    \caption{LiDAL-assisted RLNC-NOMA achievable sum rate.}
    \label{fig:LiDAL-NOMA-sumrate}
\end{figure}

Gain ratio power allocation (GRPA) \cite{marshoud_non-orthogonal_2016} is applied as a baseline power allocation scheme that assigns the optical AP \(k\) power ratio to the user \(i\) based on his imperfect CSI channel coefficient (i.e., \(h^*_i, i \in M_k\)), the approximated CSI channel coefficient (i.e., \(h^*_i + \Delta h^*_i\)) as in (\ref{eq:LiDAL-NOMA-CSI-approx}) and both are compared to his perfect CSI channel coefficient (i.e., \(h_i, i \in M_k\)) at all possible locations within a designated group \(k\). In Fig.\ref{fig:LiDAL-NOMA-sumrate}, the empirical CDF graphs depicted reflect the sum rate of possible locations in which user \(i\) can be located under the MIMO-LiDAL minimum resolution constraint and within the LiDAL-based groups formed as shown in Fig.\ref{fig:LiDAL-NOMA-Clusters}. 
The sum rate metric is selected to analyze how the user location information exploited affects the estimation of the CSI channel coefficients. It also evaluates the overall performance of the RLNC-NOMA system using the proposed LiDAL-assisted association scheme in the indoor environment considered. The graphs reflect that the use of location-based information provided by the MIMO-LiDAL system significantly improves the estimation of the CSI channel coefficients by 44\% on average, showing a performance comparable to the perfect CSI channel coefficients in terms of the sum rate for locations errors below \(4\) \(cm\). Here, the power ratios are assigned to the sorted user locations within each group relative to the highest estimated CSI channel coefficient w/o LiDAL assistance in that group. However, we observed that the sum rate achieved is not optimal under the selected power allocation scheme (i.e. GRPA) compared to the perfect CSI scenario.

\section{Conclusion}
{
In this work, we integrated a MIMO-LiDAL system into an RLNC-NOMA OWC network to jointly support communication and localization. We introduced an RLNC-NOMA transmission scheme that improves NOMA performance in terms of successful decoding probability and system sum rate. Distinct from LiDAL-only reference design, we exploit LiDAL-derived location information to enable low-complexity user association and to improve channel state information (CSI) estimation in realistic indoor scenarios with noisy and/or correlated CSI. We also optimized the localization accuracy of the baseline MIMO-LiDAL architecture and derived the CRLB as a benchmark for an unbiased localization estimator.}
{
Simulation results show that RLNC-NOMA outperforms conventional NOMA and OMA schemes in packet decoding reliability. Moreover, the proposed LiDAL-assisted design significantly improves CSI coefficient estimation, approaching the performance of systems with perfect CSI in the evaluated scenarios.}


%

%


\section*{Acknowledgment}
The authors would like to acknowledge funding from the Engineering and Physical Sciences Research Council (EPSRC) {TOWS (EP/S016570/2) and TITAN (EP/X04047X/2)} projects. For the purpose of open access, the authors have applied a Creative Commons Attribution (CC BY) license to any author accepted manuscript version arising. All data are provided in full in the results section of this article.

\ifCLASSOPTIONcaptionsoff
  \newpage
\fi



\bibliographystyle{IEEEtran}
\bibliography{References/references}
%

\end{document}